\documentclass[11pt]{amsart}
\usepackage{setspace}
\usepackage[english]{babel}
\usepackage{amsmath, amsthm}
\usepackage{amsfonts}
\usepackage{amsaddr}
\usepackage{mathtools}
\usepackage{amssymb}
\usepackage{verbatim} 
\usepackage{float}
\usepackage[pdftex]{graphicx}
\usepackage{epstopdf}
\usepackage{float}
\usepackage{caption}
\usepackage{subcaption}
\usepackage{bm}
\usepackage{dsfont}
\usepackage[hyperref]{xcolor}
\usepackage{tikz}
\usepackage[margin=1.3in]{geometry}
\usepackage{pgfplots}
\usetikzlibrary{shapes.geometric}
\usetikzlibrary{plotmarks}
\usetikzlibrary{positioning}
\usepackage{url}
\usepackage[shortlabels]{enumitem}
\usepackage{hyperref}
\usepackage{cleveref}
\crefname{figure}{Figure}{Figures}
\crefname{exmp}{example}{examples}
\crefname{equation}{equation}{equations}
\crefname{appendix}{appendix}{appendices}
\usepackage[utf8]{inputenc} %
\usepackage{multirow}

\usepackage{csquotes} %

\usepackage[
	sortcites,
	style=nature,
	giveninits=true,%
	backend=biber,
	doi=false,
	isbn=false,
	url=true,
	citestyle=numeric,
	urldate=long]{biblatex}	
\AtEveryBibitem{%
  \clearfield{month}
  \ifentrytype{book}{%
  \clearfield{pages}%
  }{%
  }%
  \ifentrytype{misc}{%
  }{%
    \clearlist{language}
    \clearfield{url}%
    \clearfield{urldate}%
  }%
}

\usepackage{algorithm}
\usepackage{enumerate}
\usepackage{algpseudocode}
\newcommand*\Let[2]{\State #1 $\gets$ #2}
\algrenewcommand\algorithmicrequire{\textbf{Input:}}
\let\oldReturn\Return
\renewcommand{\Return}{\State\oldReturn}
\newcommand{\Break}{\State \textbf{break} }

\newcommand{\oxblue}[1]{\textcolor{oxfordblue}{#1}}
\newcommand{\F}{\ensuremath{\mathbb{F}}}
\newcommand{\N}{\ensuremath{\mathbb{N}}}
\newcommand{\Z}{\ensuremath{\mathbb{Z}}}
\newcommand{\Q}{\ensuremath{\mathbb{Q}}}
\newcommand{\R}{\ensuremath{\mathbb{R}}}
\newcommand{\C}{\ensuremath{\mathbb{C}}}
\newcommand{\s}{\ensuremath{\mathbb{S}}}
\renewcommand{\H}{\ensuremath{\mathcal{H}}}
\newcommand{\D}{\ensuremath{\mathcal{D}}}
\newcommand{\T}{\ensuremath{\mathcal{T}}}
\newcommand{\A}{\ensuremath{\mathcal{A}}}
\newcommand{\M}{\ensuremath{\mathcal{M}}}
\newcommand{\K}{\ensuremath{\mathcal{K}}}
\newcommand{\vect}[1]{\boldsymbol{\mathbf{#1}}}
\newcommand{\id}{\,\mathrm{d}}
\newcommand{\Ex}[1]{\ensuremath{\mathbb{E}\left[ #1 \right]}}

\newtheorem*{lemma*}{Lemma}

\theoremstyle{definition}

\definecolor{blue}{rgb}{0.2980392156862745, 0.4470588235294118, 0.6901960784313725}
\definecolor{green}{rgb}{0.3333333333333333, 0.6588235294117647, 0.40784313725490196}
\definecolor{red}{rgb}{0.7686274509803922, 0.3058823529411765, 0.3215686274509804}
\definecolor{purple}{rgb}{0.5058823529411764, 0.4470588235294118, 0.6980392156862745}
\definecolor{yellow}{rgb}{0.8, 0.7254901960784313, 0.4549019607843137}
\definecolor{cyan}{rgb}{0.39215686274509803, 0.7098039215686275, 0.803921568627451}
\definecolor{bb1}{rgb}{0.0,  0.44705882,  0.69803922}
\definecolor{bb2}{rgb}{0.0,  0.61960784,  0.45098039}
\definecolor{bb3}{rgb}{0.83529412,  0.36862745,  0.0}
\definecolor{bb4}{rgb}{0.8,  0.4745098 ,  0.65490196}
\definecolor{bb5}{rgb}{0.94117647,  0.89411765,  0.25882353}
\definecolor{bb6}{rgb}{0.3372549 ,  0.70588235,  0.91372549}
\definecolor{orange}{rgb}{0.83529412,  0.36862745,  0.0}
\definecolor{oxfordblue}{rgb}{0.00000,  0.12941,  0.27843}

\graphicspath{{Graphics/}} 
\newlength\figureheight 
\newlength\figurewidth 
\hypersetup{
    linkcolor = oxfordblue,
    anchorcolor = oxfordblue,
    citecolor = oxfordblue,
    filecolor = oxfordblue,
    urlcolor = oxfordblue,
    colorlinks  = true,
}

\title[Uniformisation of chemical reaction networks]{Uniformisation techniques for stochastic simulation of chemical reaction networks}
\author[ C.~H.~L.~Beentjes and R.~E.~Baker]{Casper H.~L.~Beentjes  and Ruth~E.~Baker}
\address{Mathematical Institute, University of Oxford, Oxford, UK}
\email{\href{mailto:beentjes@maths.ox.ac.uk}{beentjes@maths.ox.ac.uk}}
\thanks{\oxblue{This article may be downloaded for personal use only. Any other use requires prior permission of the author and AIP Publishing. This article appeared in C.~H.~L.~Beentjes and R.~E.~Baker, J. Chem. Phys. \textbf{150}, 154107 (2019) and may be found at \href{https://doi.org/10.1063/1.5081043}{https://doi.org/10.1063/1.5081043}.}}

\begin{document}

\begin{abstract}
This work considers the method of uniformisation for continuous-time Markov chains in the context of chemical reaction networks. Previous work in the literature has shown that uniformisation can be beneficial in the context of time-inhomogeneous models, such as chemical reaction networks incorporating extrinsic noise. This paper lays focus on the understanding of uniformisation from the viewpoint of sample paths of chemical reaction networks. In particular, an efficient pathwise stochastic simulation algorithm for time-homogeneous models is presented which is complexity-wise equal to Gillespie's direct method. This new approach therefore enlarges the class of problems for which the uniformisation approach forms a computationally attractive choice. Furthermore, as a new application of the uniformisation method, we provide a novel variance reduction method for (raw) moment estimators of chemical reaction networks based upon the combination of stratification and uniformisation.

\end{abstract}

\maketitle
\section{Introduction}
Mathematical modelling has become an indispensable tool in the study of chemical reaction networks. Models often include stochastic effects to account for the intrinsic and extrinsic noise of the chemical system under study. This presence of noise in cellular reaction networks has been observed experimentally \cite{McAdams1999,Elowitz2002,Blake2003,Cai2006} and the addition of stochastic effects can lead to phenomena such as stochastic focusing \cite{Paulsson2000} or resonance-inducing oscillations \cite{Hou2003}, that are not observed in the deterministic counterpart models.

Commonly, stochastic models take the form of a continuous-time Markov chain (CTMC) and whilst many such models are often deceivingly simple to write down, analytic exploration is only possible for a marginal class of models \cite{Jahnke2006,Reis2018}. In order to study these stochastic models one therefore frequently has to rely on either approximate methods  \cite{MacNamara2008,Munsky2006,Hegland2007}, which can work well for low-dimensional problems, or on stochastic simulation if the problem dimension is too high for the approximate methods to be feasible. Stochastic simulation involves generating realisations of the process underlying a model by use of a stochastic simulation algorithm (SSA). The sample paths generated by these SSAs are then used to calculate Monte Carlo (MC) summary statistics to interrogate the system dynamics. These MC methods, however, suffer from an important drawback, namely the convergence rate is of the order $N^{-1/2}$, where $N$ is the number of MC trajectories simulated. This slow convergence can render the exploration of models with high variance in copy numbers computationally intractable with existing simulation methods.

In order to enlarge the class of problems that can be studied computationally work has therefore focussed on either ``speeding up'' the SSAs in order to generate more sample paths in a given time span, or on reducing the variance of the summary statistics employed. The former approach has been relatively successful and has lead to a range of variations on the gold standard Gillespie's direct method (DM) \cite{Gillespie1977,Anderson2007,Gibson2000,Cao2004a,McCollum2006,Yates2013} that are widely employed. Variance reduction techniques, on the other hand, have traditionally received less attention in the field of stochastic simulation of chemical reaction networks. However, with the recent advent of multi-level Monte Carlo (MLMC) \cite{Anderson2012,BenHammouda2016,Lester2016} and quasi-Monte Carlo (QMC) methods \cite{Hellander2008,Beentjes2018} it was shown that it is possible to use variance reduction ideas from related fields in the context of chemical reaction modelling, which can lead to large computational gains.

In this work we focus on the application of uniformisation of CTMCs to the stochastic simulation of chemical reaction networks. This method has been used previously in different forms in \cite{Voliotis2016,Anderson2018,Hellander2008,Zhang2010} and we show here how one can improve the computational efficiency of SSAs based on uniformisation. This makes stochastic simulation with uniformisation competitive with the standard SSAs currently available. In addition to this, we also provide a new application of the uniformisation technique; we combine uniformisation with stratification to create a new variance reduction technique, which, importantly, can be applied at no extra computational cost. 

\subsection{Outline}
This paper is structured as follows. In \Cref{sec:stoch-sim} we review the basic aspects of stochastic modelling and simulation of chemical reaction networks. \Cref{sec:uniformisation} introduces the uniformisation method and shows how it can be applied to the stochastic simulation of chemical reaction networks in an efficient way. Building on the uniformisation method, \Cref{sec:stratification} introduces a new variance reduction method combining uniformisation and stratification. The applicability of this new application of uniformisation is demonstrated by examples. Finally, the concluding remarks can be found in \Cref{sec:discussion}.

\section{Stochastic simulation}\label{sec:stoch-sim}
In this work we consider stochastic simulation of a model describing the temporal evolution of molecule copy numbers in a well-mixed system of volume $V$, i.e.\ we ignore any spatial information and assume that the molecules are homogeneously distributed within $V$. There exists a natural extension of this model that includes spatial information and we refer to \cite{Erban2007, Gillespie2013} for more detail on how to include the spatial movement of molecules in the framework described here.

We consider a collection of $n$ types of chemical species $X_1,\dots,X_n$ and define $\vect{X}(t)$ to be the state vector, describing the evolution of all the species as time evolves, i.e.\ $\vect{X}_i(t)$ is the number of molecules $X_i$ at time $t$. We assume the system starts with $\vect{X}(0)=\vect{X}_0$ molecules, though it is also possible to describe an initial distribution of particles. These species $X_i$ can interact via $K$ different types of reactions $R_1,\dots,R_K$, referred to as reaction channels. In general form, such a reaction $R_k$ can be written as

\begin{equation*}
\alpha_{1,k} X_1 + \dots + \alpha_{n,k} X_n \xrightarrow{c_k} \beta_{1,k} X_1 + \dots + \beta_{n,k} X_n,\qquad k=1,\dots,K,
\end{equation*}
where $\alpha_{i,k},\beta_{i,k}\in \N$ and $c_k$ is the reaction rate constant for reaction channel $R_k$. For each reaction $R_k$ we define the stoichiometric vector $\vect{\zeta} _k$, which  defines the change in the copy number when reaction $R_k$ fires, i.e.\ the $i$-th component of $\vect{\zeta}_k$ equals $\beta_{i,k}-\alpha_{i,k}$. If the system is in state $\vect{X}(t)$ and reaction $R_k$ fires at $t+\tau$ the system jumps to the new state $\vect{X}(t + \tau) = \vect{X}(t) + \vect{\zeta}_k$.

The temporal evolution of the chemical species can now be described as
\begin{equation}
\vect{X}(t) = \vect{X}_0 + \sum_{k=1}^K N_k(t) \vect{\zeta}_k,
\end{equation}
where $N_k(t)$ denotes the number of times that reaction channel $R_k$ fires in the time interval $[0,t)$. We associate every reaction channel, $R_k$, with a propensity function, $a_k(\vect{X}(t))$, which describes the probability that the reaction channel fires in the infinitesimal time interval $[t,t+\id t)$ as follows
\begin{equation}
\mathbb{P}\left(R_k\text{ fires in }[t,t+\id t)\right) = a_k(\vect{X}(t))\id t.
\end{equation}
We also define the total reaction propensity $a_0(\vect{X}(t)) = \sum_{k=1}^K a_k(\vect{X}(t))$. For this work we primarily use \textit{mass action kinetics}, under which the propensity for reaction $R_k$ to fire is proportional to the number of possible combinations of reactants, see for example \cite{Higham2008}. However, note that the selection of this function is a modelling choice and not essential for anything that follows.

The modelling choice for the counting functions, $N_k(t)$ for $k=1,\dots,K$, that we follow here is such that the model describing the evolution of $\vect{X}(t)$ is a CTMC. This can be described by the Kurtz random time change representation (RTCR)\cite{Anderson2011}, where the number of times reaction $R_k$ fires is described by an inhomogeneous Poisson counting process $Y_k$ so that
\begin{equation}\label{eq:KurtzRT}
\vect{X}(t) = \vect{X}_0 + \sum_{k=1}^K Y_k\left(\int_0^t a_k(\vect{X}(s)) \id s\right) \vect{\zeta}_k.
\end{equation}
Here the counting processes $Y_k$ for $k=1,\dots,K$, are all independent, reflecting the independence of the firing of reactions. The RTCR \eqref{eq:KurtzRT} yields a pathwise representation of the dynamics of the species in the system. An alternative description of the same dynamics is by use of the chemical master equation (CME), which comprises a (high-dimensional) system of ordinary differential equations (ODEs). The CME fully describes the evolution in time of the distribution of the occupation probability of the state space of $\vect{X}(t)$. Only for a small class of problems does the CME have a known analytical solution \cite{Jahnke2006,Reis2018} and one therefore often has to rely on approximate methods or stochastic simulation via Monte Carlo methods to explore the behaviour of the system. Here we focus on the stochastic simulation approach and we refer to \cite{Schnoerr2017} for an overview of alternative methods that build on (computational) approximations to the CME, such as the finite state projection method \cite{MacNamara2008a,Gupta2017,Munsky2006}, moment-closure methods \cite{Bronstein2018,Schnoerr2015} and system-size expansions \cite{VanKampen2007}.

One of the most widely used methods to generate exact realisations of the RTCR \eqref{eq:KurtzRT} is Gillespie's DM \cite{Gillespie1977}, as depicted in \Cref{algo:SSA}. The DM constructs sample trajectories which are sampled from the exact distribution described by the CME.

\begin{algorithm}
\caption{Gillespie's direct method (DM)}
\label{algo:SSA}
\begin{algorithmic}[1]
\Require{Initial data $\vect{X}_0$}
\Require{Stoichiometric matrix $\vect{\zeta}$}
\Require{Propensity functions $a_k(\vect{x})$}
\Require{Final time $T$}

\Let{$\vect{X}$}{$\vect{X}_0$}
\Let{$t$}{0}
\While{$t<T$}
\State{Generate $u_1,u_2 \sim \mathcal{U}(0,1)$}
\Let{$\vect{A}_k$}{$a_k(\vect{X})$} \Comment{Calculate the propensities.}
\Let{$a_0$}{$\sum_k \vect{A}_k$} \Comment{Calculate the total reaction propensity.}
\Let{$\tau$}{$-\log u_1/a_0$} \Comment{Calculate the next reaction time.}
\Let{$t$}{$t+\tau$}
\If {$t\geq T$}
\Let{$t$}{$T$}
\Break
\EndIf
\State{Find $p$ such that $\sum_{k=1}^{p-1}\vect{A}_k<a_0 u_2\leq \sum_{k=1}^{p}\vect{A}_k$} \Comment{Choose next reaction to fire.}
\Let{$\vect{X}$}{$\vect{X}+\vect{\zeta}_p$} \Comment{Update state vector.}
\EndWhile
\Return{$\vect{X}$}
\end{algorithmic}
\end{algorithm}

Given an ensemble of simulated sample paths $\vect{X}^{(i)}(t)$ for $i=1,\dots,N$, we want to estimate some desired summary statistic of the system of interest
\begin{equation}
Q = \Ex{f\left(\vect{X}(t)\right)}.
\end{equation}
This can be achieved by the usual Monte Carlo estimate combining the sample paths
\begin{equation}\label{eq:MC-summarystatistic}
\hat{Q} = \frac{1}{N}\sum_{i=1}^N f\left(\vect{X}^{(i)}(t)\right).
\end{equation}
The individual realisations of $\hat{Q}$ will in general not be equal to $Q$ and are random variates which therefore have an inherent uncertainty related to them. This uncertainty can be quantified by the mean squared error (MSE)
\begin{equation}
\text{MSE}\left[\hat{Q}\right] = \mathbb{V}\left[\hat{Q}\right] + \text{Bias}^2\left[\hat{Q}\right].
\end{equation}
The sample paths from the DM are unbiased and the MSE is therefore purely determined by the variance of the estimator $\hat{Q}$. Given a sample variance $\sigma^2$ of the desired summary statistic we know that $\mathbb{V}[\hat{Q}] = \sigma^2/N$ with $N$ the number of sample paths \cite[Chapter 1]{Lemieux2009}.

For a fixed amount of computational resource we would like to have a MSE that is as small as possible. A drawback of the DM is that it generates each and every reaction in the system individually, which can incur large computational costs for systems with many particles and/or many reactions, or that run for a long time. In that particular case the computational budget is only sufficient for a low number of sample paths, $N$, resulting in a high variance of the estimator $\hat{Q}$. There exists a variety of (computational) modifications to the DM \cite{Anderson2007,Gibson2000,Cao2004a,McCollum2006,Yates2013} which aim to improve the runtime by optimising certain steps, predominantly steps 5 and 8, of \Cref{algo:SSA} without affecting the statistically exact sampling. These methods therefore result in a smaller MSE for a given computational budget by increasing the number of sample paths, $N$, that can be generated in fixed time. An alternative approach is to allow the sampling to be only approximately correct in order to speed up computations, and this yields methods such as Gillespie's $\tau$-leaping \cite{Gillespie2001} and $R$-leaping \cite{Auger2006}, which introduce a bias into the MSE by virtue of only being approximate methods. A computational complexity analysis of some of the standard SSAs such as Gillespie's DM and $\tau$-leaping in the classical population scaling can be found in \cite{Anderson2018a}.

As an alternative approach we can attempt to generate sample paths using a different method which yields a smaller sample variance, $\sigma^2$. This class of techniques is known as variance reduction methods \cite[Chapter 4]{Lemieux2009}. In the context of the CME, several variance reduction methods exist, such as QMC methods \cite{Hellander2008,Beentjes2018} and MLMC methods \cite{Anderson2012,BenHammouda2016,Lester2016}. In this work we will demonstrate how the application of the uniformisation technique can lead to a new variance reduction method for the simulation of chemical reaction networks.

\section{Uniformisation}\label{sec:uniformisation}
The uniformisation method is a well-known method in probability theory, and it can be used to convert a CTMC into a discrete-time Markov chain (DTMC). For an overview of uniformisation from a probability perspective we refer to \cite{VanDijk2018}. Here, however, we describe how to employ the uniformisation technique in the context of the simulation of exact sample paths according to the RTCR, which has previously appeared, albeit in slightly different forms and with different motivation, in \cite{Voliotis2016,Anderson2018,Hellander2008,Zhang2010,Thanh2018,Sandmann2008}. For simplicity we make the assumption that the reaction propensities, $a_k$, have no explicit time-dependence and only depend on time through the state vector $\vect{X}(t)$. This assumption restricts the class of problems to time-homogeneous Markov chains and thus excludes, for example, time-dependent reaction rates. It is, however, possible to extend most of the results in this work by relaxing this condition, as we show in the \Cref{ap:time-inhomogeneous}.

We start with the original system from the \Cref{sec:stoch-sim} containing $K$ reaction channels and consider an extension of this system by adding a new reaction channel $R_{K+1}$ which takes the trivial form
\begin{equation}
\emptyset \xrightarrow{} \emptyset,
\end{equation}
and we denote this as a virtual reaction. It should be clear that the addition of this virtual reaction does not change the dynamics of $\vect{X}$, because the new channel has none of the species as reactant or product. We note that, as a result, the statistics from this new extended system are equal to that of the original system and therefore sample paths for the extended system are exact realisations of the original RTCR \eqref{eq:KurtzRT}. This observation is independent of the reaction propensity $a_{K+1}(\vect{X}(t))$ that we choose for the virtual reaction. Given some $\bar{a}>0$ we therefore use this liberty to set
\begin{equation}\label{eq:virtual-propensity}
a_{K+1}(\vect{X}(t)) = \bar{a} - a_0(\vect{X}(t)) = \bar{a} - \sum_{k=1}^K a_k(\vect{X}(t)).
\end{equation}
In order for \eqref{eq:virtual-propensity} to constitute a well-defined reaction propensity we need $\bar{a}\geq a_0(t)$ for all $t\in[0,T)$, the time interval of interest. We will come back to discuss this assumption later in this section.

The choice \eqref{eq:virtual-propensity} for the reaction propensity of the reaction channel $R_{K+1}$ might seem peculiar at first, but we note that in the new extended system the total propensity of a reaction happening is given by $a_0 + a_{K+1} = \bar{a}$, which is therefore independent of the particular state $\vect{X}(t)$ the system is in. This is in contrast to the original system where the total reaction propensity, $a_0$, is generally state dependent. We therefore end up with a uniform total reaction propensity $\bar{a}$ which we will call the uniformisation rate, and we will denote the extended system consisting of reactions $R_1,\dots,R_K,R_{K+1}$ from now on as the uniformised system. This uniformisation of the system has a few implications which we will discuss next.

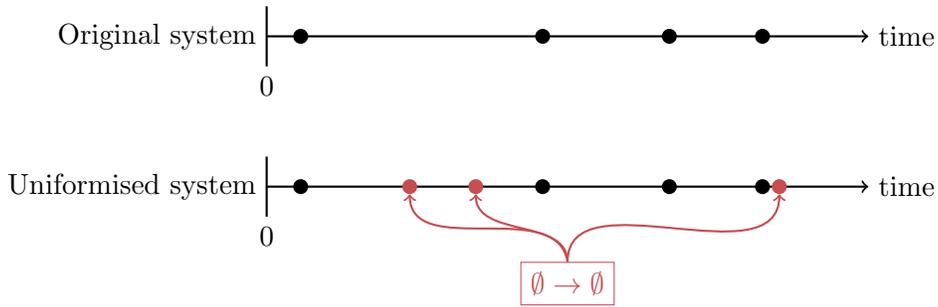
\begin{figure}[!htb]
\begin{tikzpicture}
\node[left] at (0,0) {Original system};
\draw[->, thick] (0,0) -- (8,0);
\draw[thick] (0,-0.4) -- (0,0.4);
\node[below] at (0,-0.4) {0};
\node[right] at (8,0) {time};
\node[circle,fill,inner sep=2pt] at (0.4526,0) {};
\node[circle,fill,inner sep=2pt] at (3.6689,0) {};
\node[circle,fill,inner sep=2pt] at (5.3533,0) {};
\node[circle,fill,inner sep=2pt] at (6.5915,0) {};
\node[red,below] at (4.1,-0.4) {};
\node[red,above] at (3.7,0.4) {};

\node[left] at (0,-2) {Uniformised system};
\draw[->, thick] (0,-2) -- (8,-2);
\draw[thick] (0,-2-0.4) -- (0,-2+0.4);
\node[below] at (0,-2.4) {0};
\node[right] at (8,-2) {time};
\node[circle,fill,inner sep=2pt] at (0.4526,-2) {};
\node[circle,fill,inner sep=2pt] at (3.6689,-2) {};
\node[circle,fill,inner sep=2pt] at (5.3533,-2) {};
\node[circle,fill,inner sep=2pt] at (6.5915,-2) {};
\node[red,below] at (4.1,-0.4) {};
\node[red,above] at (3.7,0.4) {};
\node[red,circle,fill,inner sep=2pt] at (1.9024,-2) {};
\node[red,circle,fill,inner sep=2pt] at (2.7802,-2) {};
\node[red,circle,fill,inner sep=2pt] at (6.8158,-2) {};
\node[red,below,rectangle,draw=red] at (4,-3) {$\emptyset\rightarrow\emptyset$};
\draw[thick,red,->] (4,-3) to[out=90, in=-90, looseness=1] (6.8158,-2.1);
\draw[thick,red,->] (4,-3) to[out=90, in=-90, looseness=1]  (2.7802,-2.1);
\draw[thick,red,->] (4,-3) to[out=99, in=-90, looseness=1]  (1.9024,-2.1);

\end{tikzpicture}
\caption{Illustration of reactions $R_1,\dots,R_K$ firing in the original system and the uniformised/extended system, indicated by dots ($\bullet$). We observe extra virtual reactions firing (\textcolor{red}{$\bullet$}) in the uniformised system.}
\label{fig:uniform-illustration}
\end{figure}

Firstly, in the uniformised system (trivially) at least as many reactions fire as in the original system, because of the addition of the independent virtual reaction channel, as illustrated in \cref{fig:uniform-illustration}. Na\"ive application of the DM to the uniformised system means we explicitly simulate every virtual reaction firing in addition to the real reactions present in the original system. As a result of the extra time taken to simulate the virtual reaction channel firing, such na\"ive application of the DM to the uniformised system yields slower run-times for exactly the same level of statistical accuracy compared to the DM for the original system.

Secondly, the reaction times in the uniformised system are independent and identically distributed (i.i.d.)\ exponential random variables with parameter $\bar{a}$. As a simple consequence of this we observe that the number of reactions, $M$, firing in a time interval $[t,t+\tau)$ is Poisson distributed with parameter $\bar{a}\tau$, which we will denote as $M\sim \mathcal{P}(\bar{a}\tau)$. Note that this is similar to the  $\tau$-leaping method, where one makes the approximation that the reaction propensities of the original system stay constant within a time interval $[t,t+\tau)$. For the uniformised system, however, this expression for the number of firing reactions is instead exact and leads to a uniformised version of the DM (UDM), as shown in \Cref{algo:GillespieUniform}.

\begin{algorithm}[!htb]
\caption{Uniformised direct method (UDM)}
\label{algo:GillespieUniform}
\begin{algorithmic}[1]
\Require{Initial data $\vect{X}_0$}
\Require{Stoichiometric matrix $\vect{\zeta}$}
\Require{Propensity functions $a_k(\vect{x})$}
\Require{Uniformisation rate $\bar{a}$}
\Require{Final time $T$}

\State{Generate $M\sim\mathcal{P}(\bar{a}T)$} \Comment{Total number of reactions that fire in $[0,T)$.}
\For{$m=1,\dots,M$}
\State{Generate $u_1 \sim \mathcal{U}(0,1)$}
\Let{$\vect{A}_k$}{$a_k(\vect{X})$} \Comment{Calculate real reaction propensities.}
\Let{$\vect{A}_{K+1}$}{$\bar{a}-\sum_{k=1}^K \vect{A}_k$} \Comment{Calculate virtual reaction propensity.}
\State{Find $p$ such that $\sum_{k=1}^{p-1}\vect{A}_k<\bar{a} u_1\leq \sum_{k=1}^{p}\vect{A}_k$} \Comment{Choose next reaction to fire.}
\If {$p\in\{1,\dots,K\}$} \Comment{Only need to fire real reactions.}
\Let{$\vect{X}$}{$\vect{X}+\vect{\zeta}_p$} \Comment{Update state vector.}
\EndIf
\EndFor
\end{algorithmic}
\end{algorithm}

In the UDM there is no explicit computation of the next reaction time (cf.\ step 7 in \Cref{algo:SSA}). For every reaction firing the UDM only needs to generate a single uniform random variate to determine which reaction takes place and, therefore, involves fewer computations. As a result the UDM will run faster than the DM applied to the uniformised system.  It is possible to generate the reaction times, if required, by noting that we can condition on the fact that $M$ reactions happen in $[0,T)$ with uniform rate. The reaction times are then distributed as the order statistics of $M$ uniform random variables in $[0,T)$.

We observe, however, that in the current form the UDM still suffers from two drawbacks. Firstly, the simulation will in general involve the firing of virtual reactions that do not contribute to the dynamics of the original system. This can therefore be thought of as computational waste and, as a result, the runtime of the UDM in its current form will still be comparatively larger than the runtime of the DM applied to the original system for time-homogeneous systems. Secondly, we mentioned that \eqref{eq:virtual-propensity} needs to be a well-defined propensity function, i.e.\ $\bar{a}\geq a_0(t)$ needs to hold for all $t\in[0,T)$. It is not clear \textit{a priori} whether such a uniformisation rate $\bar{a}$ exists\footnote{In the case of a system with a bounded state space it is (theoretically) possible to find a uniformisation rate by taking the maximum of the total propensity over all allowed states. Note, however, that the size of the state space, albeit finite, could be prohibitively large for such an approach to be practically feasible.} or what happens when $a_0$ becomes greater than $\bar{a}$ in the course of a simulation. These two issues are discussed next and we will show that the UDM can be adapted to be at least as fast as the DM applied to the original system.

\subsection{Firing virtual reactions}
In order to get around the issue of potentially slowing down the simulation by having to fire many virtual reactions we look at the distribution of the number of virtual reactions firing in between real reactions in the uniformised system. 

Suppose the system is in a state with propensity $a_0$ for the real reactions to fire. If the system is uniformised with rate $\bar{a}>a_0$ this means that the probability that the next reaction firing belongs to one of the $K$ real reaction channels is given by $a_0/\bar{a}$ and, equivalently, the probability a virtual reaction will fire next is $1-a_0/\bar{a}$. Note that when a virtual reaction fires none of the propensities of the real reactions change because none of the species $X_i$ are changed in the reaction. The repeated firing of the virtual reaction channel before a real reaction fires can therefore be viewed as a series of Bernoulli trials with probability $a_0/\bar{a}$ of success (firing a real reaction) and $1-a_0/\bar{a}$ of failure (firing a virtual reaction). In this scenario we are interested in the number of failures until the first success, i.e.\ the number of consecutive virtual reactions firing before a real reaction fires. This quantity has a well-known distribution, namely the geometric distribution, so that we have
\begin{equation}
\mathbb{P}\left(r \text{ consecutive virtual reactions before next real reaction fires} \right) = \left(1-\frac{a_0}{\bar{a}}\right)^r\frac{a_0}{\bar{a}}.
\end{equation}
As a result it is possible to fire all consecutive virtual reactions at once by sampling a single geometric random variable. This can be done efficiently by generating a uniform random variable $u\sim\mathcal{U}(0,1)$ and calculating $\lfloor \ln(u)/\ln(1-a_0/\bar{a}) \rfloor$, akin to the sampling of an exponential random variable. This observation leads to an improved version of the UDM, as depicted in \Cref{algo:IUDM}.

\begin{algorithm}[!htb]
\caption{Improved uniformised direct method}
\label{algo:IUDM}
\begin{algorithmic}[1]
\Require{Initial data $\vect{X}_0$}
\Require{Stoichiometric matrix $\vect{\zeta}$}
\Require{Propensity functions $a_k(\vect{x})$}
\Require{Uniformisation rate $\bar{a}$}
\Require{Final time $T$}

\State{Generate $M\sim\mathcal{P}(\bar{a}T)$.} \Comment{Total number of reactions that fire in $[0,T)$.}
\Let{$m$}{0} \Comment{Counter for number of reactions that have fired.}
\While{$m<M$}
\State{Generate $u_1 \sim \mathcal{U}(0,1)$.}
\Let{$\vect{A}_k$}{$a_k(\vect{X})$} \Comment{Calculate real reaction propensities.}
\Let{$a_0$}{$\sum_k \vect{A}_k$} \Comment{Calculate the total real reaction propensity.}
\State{Generate $m_{\text{virtual}}\sim\text{Geometric}(a_0/\bar{a})$} \Comment{\parbox[t]{.4\linewidth}{Number of virtual reactions firing consecutively.}}
\Let{$m$}{$m+m_{\text{virtual}}$}%
\If {$m\geq M$}
\Let{$m$}{$M$}
\Break
\EndIf
\State{Find $p$ such that $\sum_{k=1}^{p-1}\vect{A}_k<a_0 u_1\leq \sum_{k=1}^{p}\vect{A}_k$} \Comment{Choose next reaction to fire.}
\Let{$\vect{X}$}{$\vect{X}+\vect{\zeta}_p$}
\Let{$m$}{$m+1$}
\EndWhile
\end{algorithmic}
\end{algorithm}

\begin{figure}[!htb]
\includegraphics[scale=1]{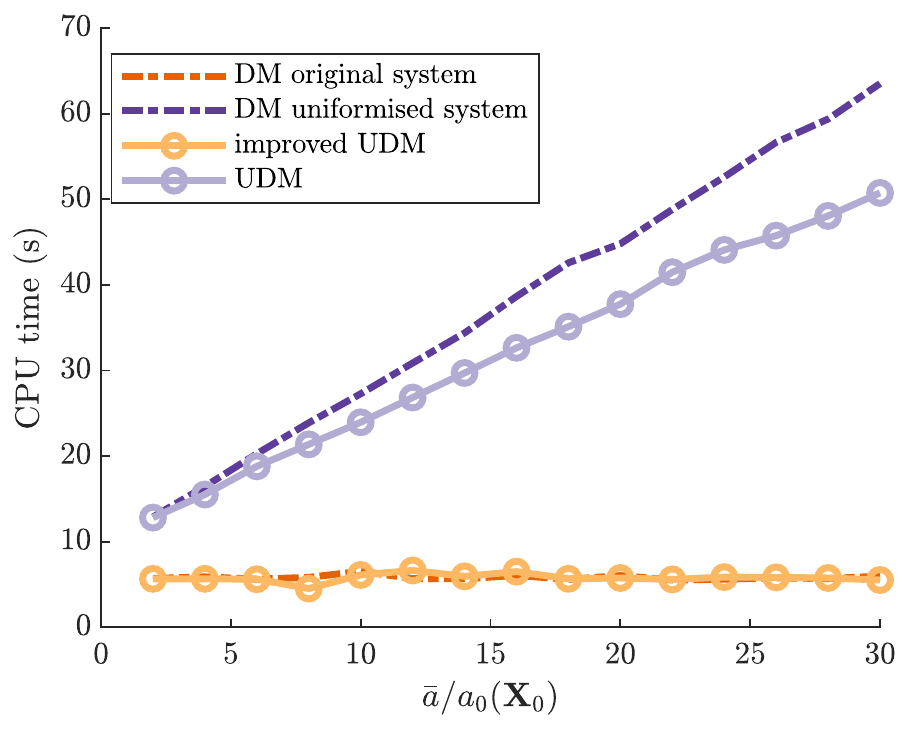}
\caption{CPU time to generate $N=10^5$ sample paths with the UDM and improved UDM compared with the DM for different uniformisation rates $\bar{a}$ relative to the initial total propensity $a_0$. Timings are for Example \ref{sec:isomerisation} with $c_1=0.2,c_2=0.1$ and $\vect{X}_0=(10,0)$, run until $T=10$. Timing experiments were performed using MATLAB R2017b on an Ubuntu desktop PC with a 3.40 GHz Intel Core i7-2600K CPU and 16 GB of random access memory.}
\label{fig:UDM-timing}
\end{figure}

In this form the improved UDM is comparable with the DM applied to the original system, i.e.\ the gold standard in the field of simulation chemical reaction networks. Both methods now fire an equal number of reactions, and need two random numbers per reaction firing. In addition to this we note that many improvements that have been made to the DM, related to speed-ups in the choice of the next reaction firing and the update of the propensities such as in \cite{Gibson2000,Cao2004a,McCollum2006,Thanh2014} and/or the re-use of random numbers \cite{Yates2013}, can be equally well applied to the improved UDM. Although it is not possible to use the well-known $\tau$-leaping approach in combination with the (improved) UDM, it is trivial to apply the $R$-leap approximation \cite{Auger2006} to the (improved) UDM. The improved version of the UDM is also insensitive to the uniformisation rate, $\bar{a}$, as can be seen in \Cref{fig:UDM-timing}. This can be intuitively understood from the observation that increasing or decreasing $\bar{a}$ only changes the number of consecutive virtual reactions firing which is taken care of in a single step in the improved UDM. 

We note from \Cref{fig:UDM-timing} and the description of the improved UDM that the computational run-time of simulations of the uniformised system can be brought back to the speed of the DM applied to the original system. However, there is no intrinsic reason to expect the improved UDM to be faster than the DM for the original system\footnote{This observation is mainly true for time-homogeneous systems, as considered in this section. For time-inhomogeneous systems, methods based on the DM for the original system can often compare unfavourably with uniformisation-based methods, as discussed in \Cref{ap:time-inhomogeneous} and \cite{Voliotis2016}.} and this is also observed in \Cref{fig:UDM-timing}. A computational speed-up, as such, is therefore not a sufficient motivation to employ the uniformisation technique in the stochastic simulation of chemical reaction networks. However, as mentioned earlier, the usage of uniformisation allows one to consider new applications, such as variance reduction methods, that are not possible under the standard SSAs for the original system.

\subsection{Adapting the uniformisation rate}\label{sec:adaptive-uniformisation}
In the exposition so far the uniformisation rate has been a free parameter and we mentioned the caveat that, by expression \eqref{eq:virtual-propensity}, $\bar{a}\geq a_0(t)$ should hold for all $t\in[0,T)$ in order for the system to remain a well-defined chemical reaction model.  We therefore discuss here the ramifications of a simulation in which the total propensity $a_0(t)$ exceeds $\bar{a}$.

\begin{figure}[!htb]
\includegraphics[scale=1]{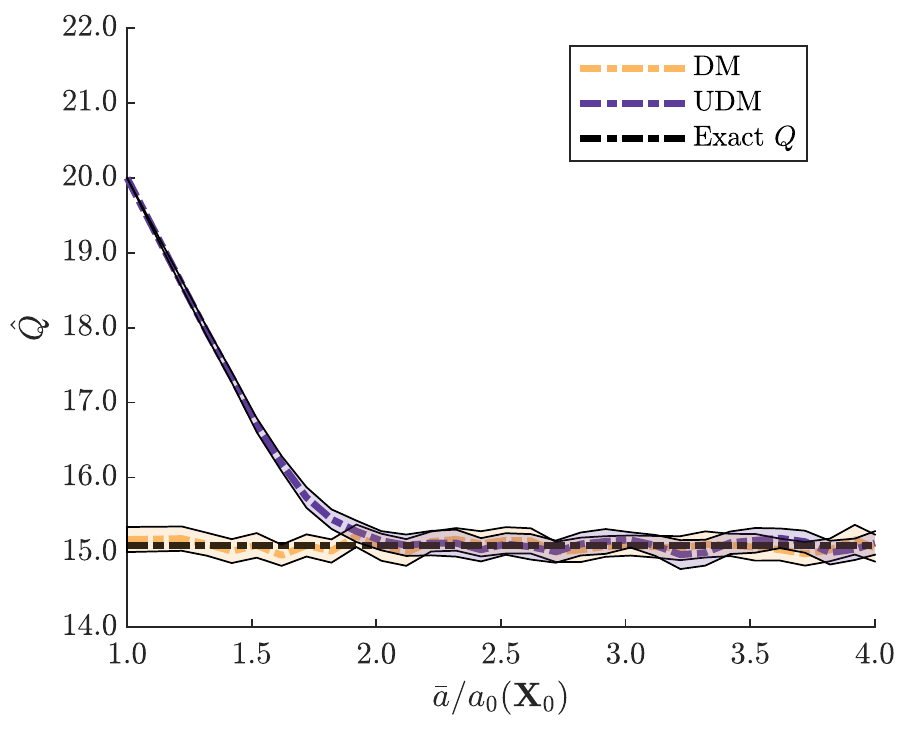}
\caption{Bias effect due to invalid samples of the uniformised system for different uniformisation rates $\bar{a}$ relative to the initial total propensity $a_0$. Model system used is Example \ref{sec:isomerisation} with $c_1=0.4,c_2=0.1$ and $\vect{X}_0=(0,20)$, run until $T=10$. The summary statistic is the number of $X_1$ molecules at final time $T$ and estimators are shown with 99.5\% confidence interval for $N=10^3$ samples.}
\label{fig:UDM-bias}
\end{figure}

A first approach to the situation $a_0>\bar{a}$ in the course of a simulation might be to classify such sample paths as invalid and generate a new, independent, sample path for each of the invalid paths until we end up with the desired number of sample paths that all satisfy $a_0\leq \bar{a}$ across $[0,T)$. This approach, unfortunately, comes at a cost, because it introduces a bias due to the new sampling strategy. Effectively, this approach only samples from a subset of the distribution defined by the CME and it is clear from \Cref{fig:UDM-bias} that the effect of the resulting bias can be dramatic and should be avoided if possible. We therefore propose a way to keep samples for which the uniformisation rate is breached, whilst keeping the estimator unbiased at the same time.

To do so, we start with a uniformised system with uniformisation rate $\bar{a}$ and sample the number of reactions $M\sim\mathcal{P}(\bar{a}T)$ firing in $[0,T)$. Now suppose that we observe that $a_0>\bar{a}$ after $M^*\leq M$ reactions have fired. Rather than discarding this sample path completely we can ask at what time $T^*\in[0,T)$ reaction $M^*$ took place, i.e.\ at what time did $a_0$ become larger than $\bar{a}$? As mentioned earlier, given that $M$ reactions fire in $[0,T)$, the reaction times $t_1,\dots,t_M$ are uniformly distributed on $[0,T)$. The reaction time $T^*=t_{M^*}$ of the $M^*$-th reaction out of $M$ reactions therefore follows the distribution of the $M^*$-th order statistic of a collection of $M$ uniform random variables on $[0,T)$. This is a well-known distribution and leads to $T^*/T\sim\text{Beta}(M^*,(M-M^*)+1)$. 

We can therefore sample $T^*$ and use the Markov property to restart the simulation from the state after $M^*$ reactions for the remaining time interval $[T^*,T)$ with a new uniformisation rate, $\bar{a}_{\text{new}}$. This procedure to describe $T^*$ conditional on $M$ reactions in $[0,T)$ leads to the correct unconditional distribution for $T^*$ and therefore does not introduce a bias, see \Cref{ap:adapting} for more detail. If we choose this new uniformisation rate such that $\bar{a}_{\text{new}}\geq a_0(T^*)$, the simulation for $[T^*,T)$ can be done in exactly the same manner with the (improved) UDM, and this approach is illustrated in \Cref{fig:adaptive-uniformisation}.

\begin{figure}[!htb]
\begin{subfigure}[b]{\textwidth}
\centering
\begin{tikzpicture}
\draw[->, thick] (0,0) -- (8,0);
\draw[thick] (0,-0.4) -- (0,0.4);
\draw[thick] (7.5,-0.4) -- (7.5,0.4);
\node[below] at (0,-0.4) {0};
\node[below] at (7.5,-0.4) {$T$};
\node[right] at (8,0) {time};
\node[below] at (2,0) {\footnotesize $\bar{a}$ total propensity};
\node[above] at (2,0) {\footnotesize $M$ reactions};
\node[red,below] at (4.1,-0.4) {};
\node[red,above] at (3.7,0.4) {};
\end{tikzpicture}
\caption{With uniformisation rate $\bar{a}$ we sample $M$ reactions to fire in $[0,T)$.}
\end{subfigure}
\\~\\
\begin{subfigure}[b]{\textwidth}
\centering
\begin{tikzpicture}
\draw[->, thick] (0,0) -- (8,0);
\draw[thick] (0,-0.4) -- (0,0.4);
\draw[thick] (7.5,-0.4) -- (7.5,0.4);
\node[below] at (0,-0.4) {0};
\node[below] at (7.5,-0.4) {$T$};
\node[right] at (8,0) {time};
\node[below] at (2,0) {\footnotesize $\bar{a}$ total propensity};
\node[red,circle, fill, inner sep =1.5pt] at (3.7,0) {};
\draw[red,thick,->] (4.1,-0.6) to[out=90, in=-90, looseness=1] (3.7,-0.1);
\node[red,below,rectangle,draw=red] at (4.1,-0.6) {e.g. $a_0>\bar{a}$};
\node[red,above] at (3.7,0.2) {$M^*$};
\end{tikzpicture}
\caption{After $M^*$ reactions we want to adapt the uniformisation rate, for example because the uniformisation rate is breached, $a_0>\bar{a}$.}
\end{subfigure}
\\~\\
\begin{subfigure}[b]{\textwidth}
\centering
\begin{tikzpicture}
\draw[->, thick] (0,0) -- (8,0);
\draw[thick] (0,-0.4) -- (0,0.4);
\draw[thick] (7.5,-0.4) -- (7.5,0.4);
\node[below] at (0,-0.4) {0};
\node[below] at (7.5,-0.4) {$T$};
\node[right] at (8,0) {time};
\node[below] at (2,0) {\footnotesize $\bar{a}$ total propensity};
\node[above] at (2,0) {\footnotesize $M^*$ reactions};
\node[red,circle, fill, inner sep =1.5pt] at (3.7,0) {};
\draw[red,thick] (3.7,-0.4) -- (3.7,0.4);
\node[red,below] at (3.7,-0.4) {$T^*$};
\node[below] at (5.5,0) {\footnotesize $\bar{a}_{\textup{new}}$ total
 propensity};
\node[above] at (5.5,0) {\footnotesize $M_{\textup{new}}$ reactions};
\end{tikzpicture}
\caption{We sample the time $T^*$ at which to adapt the uniformisation rate and restart the simulation at $T^*$ with a new uniformisation rate $\bar{a}_\text{new}$ and $M_{\text{new}}\sim\mathcal{P}(\bar{a}_{\text{new}}(T-T^*))$.}
\end{subfigure}
\caption{Illustration of the procedure adapt the uniformisation rate.}
\label{fig:adaptive-uniformisation}
\end{figure}

The approach of changing the uniformisation rate, $\bar{a}$, as described in \Cref{fig:adaptive-uniformisation}, can be applied throughout the simulation and yields a method akin to adaptive uniformisation \cite{VanMoorsel1994}. We note that the adaptation of the uniformisation rate is not restricted to time points $T^*$ at which $a_0(T^*)>\bar{a}$ is observed, adapting the uniformisation rate can be done at any point in the simulation, as shown in \Cref{algo:adaptive-IUDM}, and yields unbiased sample paths. For example, one could have a system that initially has a high total propensity, $a_0$, requiring a high uniformisation rate, $\bar{a}$, but after $M^*$ reactions $a_0$ dips considerably. In this scenario one might want to adapt the uniformisation rate, and make it smaller. Such an approach is similar to the adaptive fluctuation intervals as used in the rejection-based SSA algorithms \cite{Thanh2014,Thanh2015,Thanh2016}. An even more extreme choice might be to update the uniformisation rate after every reaction, which would yield a procedure akin to the Extrande-method \cite{Voliotis2016}.

\begin{algorithm}[!htb]
\caption{Adaptive improved uniformised direct method}
\label{algo:adaptive-IUDM}
\begin{algorithmic}[1]
\Require{Initial data $\vect{X}_0$}
\Require{Stoichiometric matrix $\vect{\zeta}$}
\Require{Propensity functions $a_k(\vect{x})$}
\Require{Uniformisation rate $\bar{a}$}
\Require{Final time $T$}

\State{Generate $M\sim\mathcal{P}(\bar{a}T)$} \Comment{Total number of reactions that fire in $[0,T)$.}
\Let{$m$}{0} \Comment{Counter for number of reactions that have fired.}
\While{$m<M$}
\State{Generate $u_1 \sim \mathcal{U}(0,1)$}
\Let{$\vect{A}_k$}{$a_k(\vect{X})$} \Comment{Calculate real reaction propensities.}
\Let{$a_0$}{$\sum_k \vect{A}_k$} \Comment{Calculate the total real reaction propensity.}
\State{Set \texttt{adapt\char`_flag}} \Comment{For example, when $a_0>\bar{a}$ we set \texttt{adapt\char`_flag}=\texttt{true}.}
\If {\texttt{adapt\char`_flag}} \Comment{Option to adapt the uniformisation rate.}
\State{Generate $T^*/T\sim\text{Beta}(m,(M-m)+1)$}
\State{Adapt the uniformisation rate $\bar{a}$}
\State{Generate $\bar{M}\sim\mathcal{P}(\bar{a}(T-T^*))$} \Comment{Total number of reactions that fire in $[T^*,T)$.}
\Let{$M$}{$m+\bar{M}$}
\EndIf

\State{Generate $m_{\text{virtual}}\sim\text{Geometric}(a_0/\bar{a})$} \Comment{\parbox[t]{.4\linewidth}{Number of virtual reactions firing consecutively.}}
\Let{$m$}{$m+m_{\text{virtual}}$}%
\If {$m\geq M$}
\Let{$m$}{$M$}
\Break
\EndIf
\State{Find $p$ such that $\sum_{k=1}^{p-1}\vect{A}_k<a_0 u_1\leq \sum_{k=1}^{p}\vect{A}_k$} \Comment{Choose next reaction to fire.}
\Let{$\vect{X}$}{$\vect{X}+\vect{\zeta}_p.$}
\Let{$m$}{$m+1$}
\EndWhile
\end{algorithmic}
\end{algorithm}

We finally mention two observations regarding the adaptive uniformisation method. Firstly, the sample paths in this framework are again exact samples according to the distribution described by the CME and can therefore be used to create unbiased summary statistics. Secondly, from the formulation of the improved UDM in the previous section we can see that the choice of $\bar{a}$ ultimately does not influence the run-time of the simulation. One might therefore be tempted to take $\bar{a}$ very large in the hope that $\bar{a}>a_0(t)$ for all $t\in[0,T)$ and we never have to adapt the uniformisation rate $\bar{a}$. However, for some choices of the reaction propensities, such as mass action kinetics, it is not always possible to select such an upper bound for $a_0$ as a uniformisation rate, $\bar{a}$, because the reaction propensities are unbounded. This is, for example, the case in a system where at least one of the species involved in mass action kinetics is unbounded itself, a situation that is often encountered. In \cite{Hellander2008} it was suggested to run a few pre-simulations to determine an empirical upper bound to $a_0$ on $[0,T)$ as a workaround if we do not know an analytical expression for an upper bound. Alternatively, one might have other motives to not choose the uniformisation rate much higher than $a_0$, an example of such a situation will be given in \Cref{sec:stratification}. To create unbiased estimators with the (improved) UDM it can therefore be necessary to employ an adaptive uniformisation method, as described in this section.

In conclusion we can create an (improved) UDM that is equally as versatile and applicable as the DM. The main benefit of uniformisation, however, lies in the fact that it opens the door to new applications that are not possible in the DM framework, as we will see next.

\section{Variance reduction via stratification}\label{sec:stratification}
Using the (improved) UDM will yield exact sample paths from the distribution defined by the CME, just like the DM. The resulting MSE of the standard Monte Carlo estimates based on these sample paths will therefore behave exactly like the MSE for the DM. In this section, however, we will show that it is possible to create a new (unbiased) estimator that has a lower sample variance, and therefore MSE, by using a stratification strategy in combination with uniformisation. 
\subsection{Stratification basics}
Suppose we are interested in estimating a summary statistic $Q=\Ex{f(\vect{Y})}$ for some process $\vect{Y}\sim p$. For example this could be a simple mean, in which case we consider $f(x)=x$, or the probability distribution evaluated at a point $y$, for which we would take $f_y(x)=\mathds{1}_{\{x=y\}}$. Rather than directly sampling from $p$ we first divide the state space $\Omega$ of $\vect{Y}$ in a set of disjoint sets such that $\Omega=\cup_{j=1}^{J}\Omega_j$, where $J$ can be infinite. These sets will be denoted as strata. We will use the notation $\omega_{\mathcal{D}_j} = \mathbb{P}\left(\vect{Y}\in \Omega_j\right)$ and define $\mathcal{D}_j$ to be the event that $\vect{Y}\in\Omega_j$. In the case that the distribution of $\vect{Y}$ is time-dependent, i.e.\ $\vect{Y}\sim p(t)$, note that $\omega_{\mathcal{D}_j}(t)$ will also be time-dependent. With this notation, and using the fact that the sets are disjoint, we can use conditional expectation and the law of total expectation to write
\begin{equation}\label{eq:total-expectation}
Q = \Ex{f\left(\vect{Y}\right)} = \sum_{j=1}^{J} \omega_{\mathcal{D}_j} \Ex{f\left(\vect{Y}\right)|\mathcal{D}_j} = \sum_{j=1}^{J} \omega_{\mathcal{D}_j} Q_{\mathcal{D}_j},
\end{equation}
where $Q_{\mathcal{D}_j}=\Ex{f\left(\vect{Y}\right)|\mathcal{D}_j}$ denotes the summary statistic conditional on the event $\mathcal{D}_j$. If the $\omega_{\mathcal{D}_j}$ are known this suggests a new way to construct an estimator of $Q$:
\begin{equation}\label{eq:strat-estimate}
\hat{Q}_{\text{str}} = \sum_{j=1}^{J} \omega_{\mathcal{D}_j} \hat{Q}_{\mathcal{D}_j},
\end{equation}
where the $\hat{Q}_{\mathcal{D}_j}$ are estimators of $Q_{\mathcal{D}_j}$. To construct these conditional estimators we denote $\vect{Y}^{(i,j)}$ to be the $i$-th sample drawn from the conditional distribution of $(\vect{Y}|\mathcal{D}_j)$ which yields
\begin{equation}\label{eq:cond-estimate}
\hat{Q}_{\mathcal{D}_j} = \frac{1}{N_j}\sum_{i=1}^{N_j} f\left(\vect{Y}^{(i,j)}\right).
\end{equation}
Note that if conditional sampling from $(\vect{Y}|\mathcal{D}_j)$ can be carried out exactly, the $\hat{Q}_{\mathcal{D}_j}$ are unbiased estimators of $Q_{\mathcal{D}_j}$ and, as a result, $\hat{Q}_{\text{str}}$ is an unbiased estimator of the summary statistic $Q$.

The main benefit of this stratified sampling approach,  however, lies in the fact that by a judicious choice of the number of samples, $N_j$, per stratum we can make sure that $\hat{Q}_{\text{str}}$ has a lower variance than the standard Monte Carlo estimator $\hat{Q}$. To see this, note that the law of total variance decomposes the variance of our summary statistic of interest, $f(\vect{Y})$, as
\begin{equation}\label{eq:total-variance}
\sigma^2 =\mathbb{V}\left[f\left(\vect{Y}\right)\right] = \Ex{\mathbb{V}\left[f\left(\vect{Y}\right)|\mathcal{D}_j\right] } + \mathbb{V}\left[\Ex{f\left(\vect{Y}\right)|\mathcal{D}_j}\right].
\end{equation}
If we introduce $\sigma_j^2$ as the conditional variance of $f(\vect{Y})$ given $\mathcal{D}_j$ we can rewrite \eqref{eq:total-variance} as
\begin{equation}\label{eq:var-decomposition}
\sigma^2 = \sum_{j=1}^J\omega_{\mathcal{D}_j} \sigma^2_j + \sum_{j=1}^J\omega_{\mathcal{D}_j}\left(Q-Q_{\mathcal{D}_j}\right)^2.
\end{equation}
As a result we see that the sample variance of the standard Monte Carlo estimator is given by
\begin{equation}\label{eq:MC-var}
\mathbb{V}\left[\hat{Q}\right]= \frac{1}{N}\sum_{j=1}^J\omega_{\mathcal{D}_j} \sigma^2_j + \frac{1}{N}\sum_{j=1}^J\omega_{\mathcal{D}_j}\left(Q-Q_{\mathcal{D}_j}\right)^2,
\end{equation}
where $N$ again denotes the number of sample paths used. For example, this would be the sample variance using samples from the DM to calculate the estimate of the summary statistic given by \eqref{eq:MC-summarystatistic}. On the other hand, for the stratified estimator we have sample variance
\begin{equation}\label{eq:Strat-var}
\mathbb{V}\left[\hat{Q}_{\text{str}}\right]= \sum_{j=1}^J\omega_{\mathcal{D}_j}^2 \frac{\sigma^2_j}{N_j}.
\end{equation}
This expression demonstrates the fact that in order to reduce the sample variance of a stratified estimator we need to carefully specify how many samples will be used per stratum, $\mathcal{D}_j$. Perhaps the simplest and most common choice is \textit{proportional allocation}, i.e.\ given a budget of $N$ samples in total we set $N_j = \omega_{\mathcal{D}_j}N$. For this choice the stratified estimator has a sample variance that is guaranteed to be at least as small as the standard Monte Carlo sample variance:
\begin{equation}
\mathbb{V}\left[\hat{Q}_{\text{prop}}\right]= \sum_{j=1}^J\omega_{\mathcal{D}_j}^2 \frac{\sigma^2_j}{N\omega_{\mathcal{D}_j}}\leq \frac{1}{N}\sum_{j=1}^J\omega_{\mathcal{D}_j} \sigma^2_j + \frac{1}{N}\sum_{j=1}^J\omega_{\mathcal{D}_j}\left(Q-Q_{\mathcal{D}_j}\right)^2=\mathbb{V}\left[\hat{Q}\right],
\end{equation}
where $\hat{Q}_{\text{prop}}$ is given by \eqref{eq:strat-estimate} and \eqref{eq:cond-estimate} with $N_j = \omega_{\mathcal{D}_j}N$. It is therefore clear that a judicious stratification strategy, such as proportional allocation, can in fact be a variance reduction technique. Other choices of sample allocation can be made such as optimal allocation and post-stratification, leading to different sample variances. We will not discuss such strategies further here, but refer the reader to \cite[Chapter 4]{Lemieux2009}.

As a final note, we provide a method for estimating the sample variance of the stratified estimator \eqref{eq:strat-estimate}, which can be used to construct confidence intervals. Given the conditional estimators as in \eqref{eq:cond-estimate}, we can write down the unbiased conditional sample variance estimator
\begin{equation}
s^2_j = \frac{1}{N_j-1}\sum_{i=1}^{N_j} \left(f\left(\vect{Y}^{(i,j)}\right)-\hat{Q}_{\mathcal{D}_j}\right)^2.
\end{equation}
An unbiased estimator of the sample variance of the stratified estimator of the summary statistic $\hat{Q}_{\text{str}}$ is then given by
\begin{equation}\label{eq:sample-var-strat}
s^2 = \sum_{j=1}^J \omega_{\mathcal{D}_j}^2\frac{s_j^2}{N_j}.
\end{equation}
\subsection{Stratification of the number of reactions}
Stratification is a technique that can be used to incorporate some exact knowledge of part of the stochastic process into the required estimator so as to lower the sample variance. In the case of chemical reaction networks we can apply stratification ideas if we first uniformise the system with some rate $\bar{a}$. We can think of the uniformised system in terms of an extended state space of the form $\vect{Y}=(\vect{X},M)$ with $\vect{X}$ the state of the species after $M$ reactions have fired. The distribution of $M$ is known \textit{a priori} (as shown in \Cref{sec:uniformisation}) and therefore we can stratify with respect to the number of reactions, $M$, that fire in the time interval of interest, $[0,T]$. Because $M\in\N_{\geq 0}$ all we need to specify is a division of the non-negative integers into disjoint sets. For this work we consider the following stratification strategy: $\mathcal{D}_j = \{M_j\leq M<M_{j+1}\}$, for some collection of non-negative integers $M_1<M_2<\dots<M_{J+1}$. This divides the state space of $\vect{Y}=(\vect{X},M)$ by $\Omega=\cup_{j=1}^J\Omega_j$ with $\Omega_j = \{(\vect{X},M)\text{ s.t. }M_j\leq M< M_{j+1}\}$, i.e.\ $\Omega_j$ represents all sample paths where the total number of reactions that have fired lies in the prescribed range $[M_j,M_{j+1})$. Note that for this choice the strata probabilities are given by a simple sum of Poisson probabilities
\begin{equation}
\omega_{\mathcal{D}_j}(t) = \sum_{m=M_j}^{M_{j+1}-1} \frac{(\bar{a}t)^m}{m!}e^{-\bar{a}t}.
\end{equation}
To construct the conditional estimators $\hat{Q} _{\mathcal{D}_j}$ we need to be able to construct sample paths conditional on $\mathcal{D}_j$. This can be easily achieved by replacing step 1 of the (improved) UDM with drawing $M$ from the truncated Poisson distribution on $[M_j,M_{j+1})$ rather than drawing $M\sim\mathcal{P}(\bar{a}T)$, as illustrated in \Cref{algo:stratification-IUDM}\footnote{Sampling from the truncated Poisson distribution can be easily carried out via the inverse transform sampling method and the truncated cumulative distribution function (CDF), $F_{\text{truncated}}(x) = (F(x)-F(M_j))/(F(M_{j+1})-F(M_j))$, where $F(x)$ is the standard Poisson CDF.}. Therefore there is no extra cost associated with this stratification strategy compared to the original (improved) UDM and, as a result, the computational cost for the stratified estimator is comparable to that of the (improved) UDM and the DM. However, at the same time it is guaranteed to have a sample variance that is at least as small as that of the standard estimator from the DM and (improved) UDM. Before applying this new estimator to some examples we mention a few caveats and observations. 

\begin{algorithm}[!htb]
\caption{Stratification with the improved uniformised direct method}
\label{algo:stratification-IUDM}
\begin{algorithmic}[1]
\Require{Initial data $\vect{X}_0$}
\Require{Stoichiometric matrix $\vect{\zeta}$}
\Require{Propensity functions $a_k(\vect{x})$}
\Require{Uniformisation rate $\bar{a}$}
\Require{Final time $T$}
\Require{Number of samples $N$}
\Require{Stratum definitions $M_1,\dots, M_{J+1}$}

\For{$j=1,\dots,J$}
\Let{$N_j$}{$\lceil \omega_{\mathcal{D}_j}(T) N \rceil$} \Comment{Proportional allocation for number of samples per stratum}
\For{$n=1,\dots,N_j$}
\State{Generate $M$ from the truncated distribution $\mathcal{P}(\bar{a}T)$ on $[M_j,M_{j+1})$.} 
\Let{$m$}{0} 
\While{$m<M$}
\State{Generate $u_1 \sim \mathcal{U}(0,1)$}
\Let{$\vect{A}_k$}{$a_k(\vect{X})$} 
\Let{$a_0$}{$\sum_k \vect{A}_k$} 
\State{Generate $m_{\text{virtual}}\sim\text{Geometric}(a_0/\bar{a})$} 
\Let{$m$}{$m+m_{\text{virtual}}$}%
\If {$m\geq M$}
\Let{$m$}{$M$}
\Break
\EndIf
\State{Find $p$ such that $\sum_{k=1}^{p-1}\vect{A}_k<a_0 u_1\leq \sum_{k=1}^{p}\vect{A}_k$} 
\Let{$\vect{X}$}{$\vect{X}+\vect{\zeta}_p.$}
\Let{$m$}{$m+1$}
\EndWhile
\EndFor
\EndFor
\end{algorithmic}
\end{algorithm}

Firstly, in order for the stratification method to work it is important that the uniformisation rate $\bar{a}$ is in fact a valid uniformisation rate over the interval of interest $[0,T)$, i.e.\ $\bar{a}\geq a_0(t)$ for all $t\in[0,T)$. If this is not the case and we are required to adapt the uniformisation rate, the samples $(\vect{X},M)$ are not all drawn from the same distribution and cannot be combined to yield the conditional estimators $\hat{Q}_{\mathcal{D}_j}$. It is therefore necessary to either know an upper bound to the total propensity \textit{a priori} or to run some pre-simulations to generate an empirical upper bound.

Secondly, the effectiveness of stratification will hinge on the choice of strata. If we use the proportional allocation strategy we see that the variance reduction benefit of stratification over standard Monte Carlo comes from the $\frac{1}{N}\sum_{j=1}^J\omega_{\mathcal{D}_j}\left(Q-Q_{\mathcal{D}_j}\right)^2$ term that is lacking in the sample variance. This term represents the inter-strata variance, i.e.\ how much the conditional summary statistics deviate from the overall summary statistic. The larger this inter-strata variance, the bigger the variance reduction gain is. Unfortunately, it is in general not possible to know how to choose the strata so as to attain a relatively large inter-strata variance. 

In choosing the strata we have to make a choice as to the number of strata. One might be tempted to take as many strata as possible as this will in theory yield the largest variance reduction. Note that theoretically we could use an infinite number of strata if we use the strata boundaries $M_j = (j-1)$ for all $j\in\N_{N\geq 0}$. However, in order to get accurate estimators for every stratum we need at least two samples per stratum, and ideally more. Therefore, increasing the number of strata will also increase the number of sample paths needed to get accurate estimates of, for example, the sample variance. It has been observed that increasing the number of strata beyond six, under some mild assumptions, yields very little extra benefit \cite[Section 5A.8]{Cochran1977}. We will show that this is also true for the simulation of chemical reaction networks in an example in the next section.

Finally there is the choice of the strata boundaries. Without knowing the joint distribution of $(\vect{X},M)$ we simply choose to use the quantiles of the distribution of $M$ to define the strata. For example, if we require four strata we use the 4-quantiles $q_1,q_2,q_3$ to define $M_1=0,M_2=q_1,M_3=q_2,M_4=q_3$ and $M_5=\infty$. Note that this choice roughly allocates an equal proportion of the total $N$ samples to each stratum because the weights $\omega_{\mathcal{D}_j}$ are equal\footnote{Due to the discrete nature of the Poisson distribution it is not possible to define strata that exactly have equal weights.}. An alternative choice, reminiscent of work in \cite{Hellander2008}, is to define $M_1$ and $M_{J+1}$ such that $\mathbb{P}(M<M_1\text{ or }M>M_{J+1})\leq \varepsilon$. We can either use $M_j = M_1+(j-1)$ or some other stratification for $M_1\leq M\leq M_{J+1}$. This will yield a biased estimator, because the $\varepsilon$-tails of the distribution of $M$ are neglected. However, if $\varepsilon$ is sufficiently small this bias will be negligible by construction. 

\subsection{Examples}
In this part of the work we consider two examples to test the variance reduction effects of using stratification in combination with uniformisation for chemical reaction networks. The first example is a linear system involving two species and therefore it is possible to show some analytical results to complement numerical results. The second example is a MAPK cascade model comprising eight species interacting via ten reactions with non-linear reaction propensities. The results presented here are given only in terms of the variance reduction factor and not based on the relative efficiency of the methods, because this will depend heavily on the implementation details and the particular computing system used. The variance reduction factor, however, is implementation independent.

\subsubsection{Isomerisation}\label{sec:isomerisation}In this example we consider the simple reversible isomerisation of species $X_1$ into $X_2$:
\begin{subequations}\label{eq:isomerisation}
\begin{align}
X_1 &\xrightarrow{c_1} X_2;\\
X_2 &\xrightarrow{c_2} X_1.
\end{align}
\end{subequations}
Due to the linear nature of the reactions we can write down an analytical expression for the probability distribution $\mathbb{P}(\vect{X},t)$ \cite{Jahnke2006}. We note that the state space of $\vect{X}$ is bounded and as a result there exists a uniformisation rate valid for all $T>0$, e.g. $\bar{a}=\max(c_1,c_2)\cdot\left(X_1(0)+X_2(0)\right)$. Furthermore, because it is a simple linear system, we can find the joint distribution of $(X,M)$ numerically. This is achieved by writing down the transition matrix for the DTMC of the uniformised system. This allows one to numerically evaluate \eqref{eq:var-decomposition} and therefore give an analytic value for the variance reduction that can be attained. We will use the following notation to denote the variance reduction factor (VRF)
\begin{equation}
\alpha = \frac{\mathbb{V}\left[\hat{Q}\right]}{\mathbb{V}\left[\hat{Q}_{\text{str}}\right]}.
\end{equation}

\begin{figure}[!htb]
\includegraphics[scale=1]{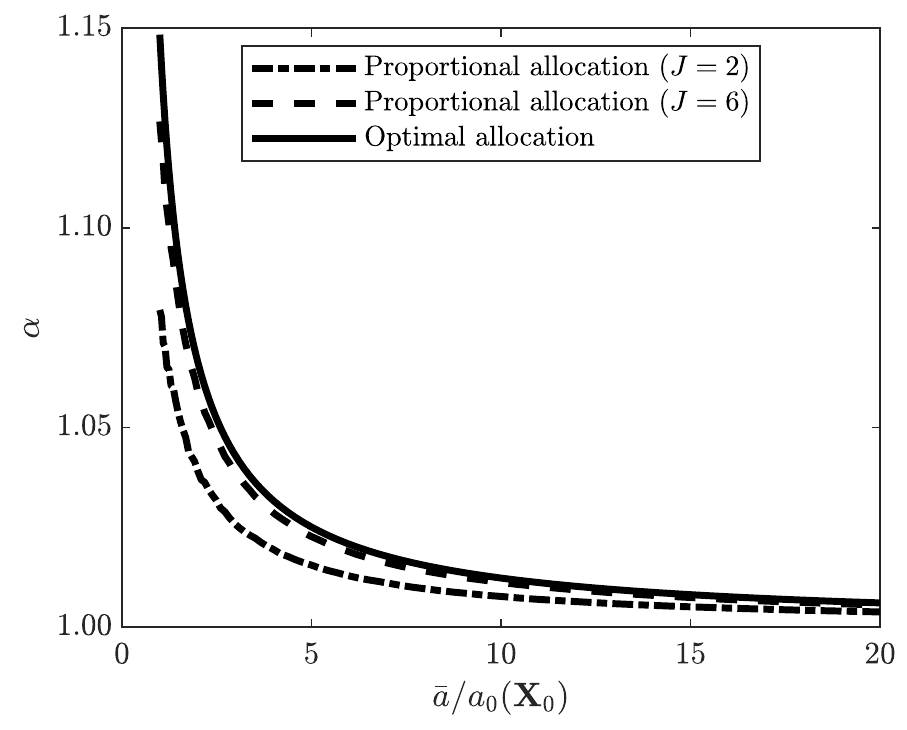}
\caption{Variance reduction factor for different uniformisation rates $\bar{a}$ relative to the initial total propensity $a_0$. Example \ref{sec:isomerisation} was run with $c_1=0.3,c_2=0.1$ and $\vect{X}_0=(20,0)$ until $T=5$. The summary statistic is the number molecules $X_1$ at final time $T$.}
\label{fig:VRF-rate}
\end{figure}

We first look at the influence of the uniformisation rate $\bar{a}$ on the VRF. Because the system is linear and effectively monomolecular we can calculate the VRF for both proportional allocation and optimal allocation. The latter method maximises the VRF, but for general systems is not a practical method. It does, however, provide an upper bound for the VRF for any stratification strategy and is therefore included here. As can be seen in \Cref{fig:VRF-rate} the VRF is moderate and decreases monotonically as a function of $\bar{a}$. Furthermore, the VRF depends on the number of strata, $J$, used and, as expected, more strata entails a larger VRF. Note, however, that as expected the VRF is always larger than unity, meaning that regardless of $\bar{a}$ there is always a (slight) variance reduction. One might be tempted to take a lower value of $\bar{a}$ than drawn in the hope of getting larger VRFs. Unfortunately, there is a balance between the uniformisation rate needing to be large enough so as to not bias the simulations and keeping $\bar{a}$ small enough to see a more significant VRF.

To see the effect of varying the number of strata we look at the same system that was used to generate \Cref{fig:VRF-rate}, but fix the uniformisation rate $\bar{a}$ whilst we vary the number of strata, $J$. The results are presented in \Cref{tab:VRF} and clearly show that increasing the number of strata beyond a moderate number like $J=6$ has only a marginal effect. We can also observe that there seems to be an inverse relationship between the uniformisation rate, $\bar{a}$, and the variance reduction factor of the form $\alpha = 1 + A(J)/\bar{a}$ for some constant $A(J)$ depending on the number of strata.

\setlength{\tabcolsep}{20pt}
\renewcommand{\arraystretch}{1.5}
\begin{table}[!htb]
\begin{tabular}{c||ccc}
\multirow{2}{*}{\begin{tabular}[c]{@{}c@{}}Number of strata\\ $J$\end{tabular}} & \multicolumn{3}{c|}{Relative uniformisation rate $\bar{a}/a_0(\vect{X}_0)$}       \\
                                                                                & 1                 & 10                & \multicolumn{1}{c|}{100}               \\ \hline \hline
2                                                                               & $7.9\cdot 10^{-2}$ & $7.7\cdot 10^{-3}$ & \multicolumn{1}{c|}{$7.7\cdot 10^{-4}$} \\
4                                                                               & $1.2\cdot 10^{-1}$ & $1.1\cdot 10^{-2}$ & \multicolumn{1}{c|}{$1.0\cdot 10^{-3}$} \\
6                                                                               & $1.3\cdot 10^{-1}$ & $1.1\cdot 10^{-2}$ & \multicolumn{1}{c|}{$1.1\cdot 10^{-3}$} \\
12                                                                              & $1.4\cdot 10^{-1}$ & $1.2\cdot 10^{-2}$ & \multicolumn{1}{c|}{$1.2\cdot 10^{-3}$} \\ \hline
$\infty$ prop.                                                                  & $1.5\cdot 10^{-1}$ & $1.2\cdot 10^{-2}$ & \multicolumn{1}{c|}{$1.2\cdot 10^{-3}$} \\
$\infty$ opt.                                                                   & $1.5\cdot 10^{-1}$ & $1.2\cdot 10^{-2}$ & \multicolumn{1}{c|}{$1.2\cdot 10^{-3}$} \\
\hline
\end{tabular}
\caption{Effect of the number of strata, $J$, and the uniformisation rate, $\bar{a}$, on the VRF. Tabulated is $\alpha-1$ (higher is better) for the isomerisation model \eqref{eq:isomerisation} with $c_1=0.3,c_2=0.1$ and $\vect{X}_0=(20,0)$, run until $T=5$. The last rows, $J=\infty$, indicate the maximal VRF that can be achieved for both proportional allocation and optimal allocation, respectively.}
\label{tab:VRF}
\end{table}

We can understand the relatively small VRF that we observe based on the decomposition of the estimator variances \eqref{eq:MC-var} and \eqref{eq:Strat-var}. The difference between the two variances is due to the inter-strata variance and it increases with an increasing inter-strata variance. However, by increasing the uniformisation rate, $\bar{a}$, the virtual reactions become more prevalent. This has the result that most of the reactions in a stratum become virtual reactions and therefore the strata distributions become more similar, diminishing the inter-strata variance.

Finally, we note that numerical verification of the VRF is computationally intensive, because the VRF is small. We used $N=2^{14}$ sample paths from the uniformised system with uniformisation rate $\bar{a}=6$ and $J=6$ strata, parameters $c_1=0.3,c_2=0.1$ and $\vect{X}_0 =(20,0)$ at time $T=5$ to find the estimator variances for the summary statistic $Q=\Ex{X_1(T)}$ using the standard Monte Carlo sample variance estimator and \eqref{eq:sample-var-strat}. By repeating this procedure 256 times, yielding roughly $4\cdot 10^6$ sample paths in total, we can construct confidence intervals for the variances of both estimators. This yields the 99.7\% confidence intervals for the standard Monte Carlo method $\mathbb{V}[\hat{Q}] \in (4.549,4.568)/N$ and for the stratified estimator $\mathbb{V}[\hat{Q}_{\text{prop}}] \in (4.035,4.052)/N$. Both of these results agree with the theoretical values of $\mathbb{V}[\hat{Q}]=4.559/N$ and $\mathbb{V}[\hat{Q}_{\text{prop}}]=4.047/N$, respectively, and show that numerically there is a variance reduction visible, albeit a small one.

\subsubsection{MAPK-cascade with feedback}\label{sec:mapk}
\begin{figure}[!htb]
\begin{center}
\scalebox{0.7}{\begin{tikzpicture}[auto, outer sep=3pt, node distance=2cm]
\tikzstyle{species}=[align=center,circle, minimum size=2.2cm,draw=gray!80,fill=gray!20]

\node[species,] (MKKK) at (0,0) {MKKK \\ $X_1$}; %
\node[species,right=of MKKK] (MKKK-P) {MKKK-P \\ $X_2$}; 
\node[species,below=of MKKK] (MKK) {MKK \\ $X_3$}; %
\node[species,below=of MKKK-P] (MKK-P) {MKK-P \\ $X_4$};
\node[species,right=of MKK-P] (MKK-PP) {MKK-PP \\ $X_5$};
\node[species,below=of MKK-P] (MAPK) {MAPK \\ $X_6$};
\node[species,below=of MKK-PP] (MAPK-P) {MAPK-P \\ $X_7$};
\node[species,right=of MAPK-P] (MAPK-PP) {MAPK-PP \\ $X_8$};

\draw[->,thick] (MKKK) to [out=45,in=135,looseness=0.7] node[below] (A) {1} (MKKK-P);
\draw[->,thick] (MKKK-P) to [out=-135,in=-45,looseness=0.7] node[above] {2} (MKKK);

\draw[->,thick] (MKK) to [out=45,in=135,looseness=0.7] node[below] (B) {3} (MKK-P);
\draw[->,thick] (MKK-P) to [out=45,in=135,looseness=0.7] node[below] (C) {4} (MKK-PP);
\draw[->,thick] (MKK-PP) to [out=-135,in=-45,looseness=0.7] node[above] {5} (MKK-P);
\draw[->,thick] (MKK-P) to [out=-135,in=-45,looseness=0.7] node[above] {6} (MKK);

\draw[->,thick] (MAPK) to [out=45,in=135,looseness=0.7] node[below] (D) {7} (MAPK-P);
\draw[->,thick] (MAPK-P) to [out=45,in=135,looseness=0.7] node[below] (E) {8} (MAPK-PP);
\draw[->,thick] (MAPK-PP) to [out=-135,in=-45,looseness=0.7] node[above] {9} (MAPK-P);
\draw[->,thick] (MAPK-P) to [out=-135,in=-45,looseness=0.7] node[above] {10} (MAPK);

\draw[->,thick,densely dashed] (MKKK-P) to [out=-90,in=90,looseness=1.5] (B);
\draw[->,thick,densely dashed] (MKKK-P) to [out=-90,in=90,looseness=1.5] (C);

\draw[->,thick,densely dashed] (MKK-PP) to [out=-90,in=90,looseness=1.5] (D);
\draw[->,thick,densely dashed] (MKK-PP) to [out=-90,in=90,looseness=1.5] (E);

\node[above=7.5 of MAPK-PP] (anchor1) {};
\node[above right =0.9cm of A] (anchor3) {};
\node[right=5cm of anchor3] (anchor2) {};

\draw[->,ultra thick,gray,dotted] (MAPK-PP) to [out=90,in=-90,looseness=1.5] (anchor1) to [out=90,in=0,looseness=1.5] (anchor2) to [out=180,in=0,looseness=1.5] (anchor3) to [out=180,in=90,looseness=1.5] (A);

\end{tikzpicture}}
\end{center}
\caption{MAPK-cascade network, adapted from \cite{Kholodenko2000}. Reaction channels are depicted by the numbered horizontal arrows. Vertical dashed lines in black and the grey dotted line denote positive and negative feedback effects, respectively.}
\label{fig:MAPK-network}
\end{figure}
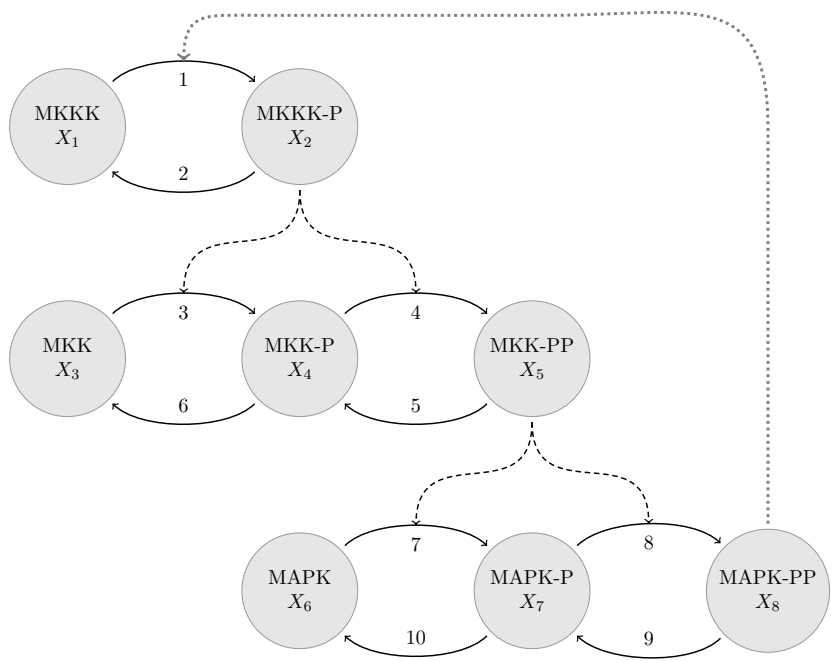

The second example is a mitogen-activated protein kinase (MAPK) cascade model from \cite{Kholodenko2000}. It consists of eight species linked by ten reaction channels with Michaelis-Menten kinetics and Hill functions. A schematic representation of the network structure is shown in \Cref{fig:MAPK-network}.

The MAPK-cascade network describes phosphorylation of MAPK by a layered process involving several kinases. MAPK, the terminal kinase, is phosphorylated by MAPK kinases (MKKs), depicted by the interaction between the second and third layer in \Cref{fig:MAPK-network}. In turn the MKKs are phosphorylated by the MAPK kinase kinases (MKKKs), which is visible as the interaction between the first and second layer in \Cref{fig:MAPK-network}. The activation of MKKK is thought to be inhibited by MAPK via membrane-bound Ras and is modelled as a negative feedback loop, as can be seen in \Cref{fig:MAPK-network}.

For this system there exist three elementary conservation laws and every chemical species is involved in one of them:
\begin{subequations}\label{eq:conservationlaw}
\begin{align}
X_1(t) +X_2(t)&=X_1(0)+X_2(0);\\
X_3(t) +X_4(t)+X_5(t)&=X_3(0)+X_4(0)+X_5(0);\\
X_6(t) +X_7(t)+X_8(t)&=X_6(0)+X_7(0)+X_8(0).
\end{align}
\end{subequations}
It is therefore possible to bound the total propensity uniformly. In this case we uniformise the system with uniformisation rate $\bar{a}=15$ and apply stratification using $J=6$ strata, i.e.\ the 6-quantiles. We use the parameter values and initial conditions ($\vect{X}_0=(100, 0, 300, 0, 0, 300, 0, 0)$) for the model as given in \cite[Table 2]{Kholodenko2000}. Because of the non-linearity of the reaction dynamics it is not possible to find analytical values for the estimator sample variances and VRF. We therefore perform 256 repeated stochastic simulations using $N=2^{14}$ sample paths per simulation to get estimates for the sample variances. As summary statistics we use the number of molecules at the final time $T=200$ s for the eight different species in the system and the results are tabulated in \Cref{tab:MAPK-var}.

\setlength{\tabcolsep}{5pt}
\renewcommand{\arraystretch}{1.5}
\begin{table}[!htb]
\begin{subtable}{1\textwidth}
\centering
\begin{tabular}{c||cccc|}
\multirow{2}{*}{Sample variance} & \multicolumn{4}{c|}{Summary statistic $Q=\mathbb{E}\left[X_i\right]$} \\
                                 & $X_1$ & $X_2$  & $X_3$  & $X_4$    \\ \hline \hline
$N\cdot\mathbb{V}\left[\hat{Q}\right]$       &   (22.6, 22.7)        &   (22.6, 22.7)    &  (202.4, 203.2)    &  (225.3, 226.2)      \\
$N\cdot\mathbb{V}\left[\hat{Q}_{\text{prop}}\right]$ &   (22.2, 22.3) &   (22.2, 22.3)     &  (193.7, 194.5)    &  (225.0, 225.9)           \\ \hline \hline
VRF & $1.016\pm 0.003$ & $1.016\pm 0.003$& $1.045\pm 0.003$ & $1.004 \pm 0.003$\\ \hline
\end{tabular}
\caption{First four species summary statistics}
\end{subtable}

\bigskip
\begin{subtable}{1\textwidth}
\centering
\begin{tabular}{c||cccc|}
\multirow{2}{*}{Sample variance}   & \multicolumn{4}{c|}{Summary statistic $Q=\mathbb{E}\left[X_i\right]$} \\
                                              & $X_5$ & $X_6$ & $X_7$ & $X_8$ \\ \hline \hline
$N\cdot\mathbb{V}\left[\hat{Q}\right]$    &  (259.6, 260.7)           &  (199.8, 201.2)     &   (347.4, 348.7)    &  (777.1, 779.9)      \\
$N\cdot\mathbb{V}\left[\hat{Q}_{\text{prop}}\right]$ &  (246.8, 247.9)  &       (180.0, 181.3)  &  (312.3, 313.6)  &  (670.6, 673.4)     \\ \hline
VRF & $1.052\pm 0.003$ & $1.110\pm 0.006$ & $1.112\pm 0.003$ & $1.159\pm 0.003$\\ \hline
\end{tabular}
\caption{Last four species summary statistics}
\end{subtable}

\caption{Sample variance and estimated VRF for the MAPK cascade for both the standard Monte Carlo estimator and the stratified estimator. System is run until $T=200$ with parameter values and initial conditions as given in \cite[Table 2]{Kholodenko2000} and using a uniformisation rate $\bar{a}=15$. Values tabulated are the 99.7\% confidence intervals.}
\label{tab:MAPK-var}
\end{table}

As we can see in \Cref{tab:MAPK-var} the VRF depends heavily on the chosen summary statistic. The MAPK-cascade is a hierarchical system in the sense that the conversion of the downstream species  $X_6,X_7$ and $X_8$ depends on other upstream species, in particular $X_5$. The inter-strata variation will therefore likely be larger. The lower quantile strata observe fewer reactions and therefore will not see much activation of the downstream species, in contrast to the higher quantile strata, where more reactions fire and we see more conversion of the downstream species. The conditional summary statistics will therefore differ significantly between the different strata, which we know leads to a larger VRF by the decomposition of the variance \eqref{eq:var-decomposition}.

We can therefore conclude that the VRF using stratification and uniformisation can vary strongly depending on the model system, the summary statistic of choice, the uniformisation rate and the number of strata used. It is, however, also true that for the same amount of computational complexity a variance at least as small as for the standard DM is observed, at no extra cost.

\section{Discussion and outlook}\label{sec:discussion}
Uniformisation techniques are well-known tools in probability theory and have previously appeared in the context of chemical reaction networks in various forms \cite{Voliotis2016,Anderson2018,Zhang2010,Hellander2008,Thanh2018}. In this work we revisited the uniformisation technique in the context of pathwise simulation of realisations of said chemical reaction networks and focussed on the question whether pathwise simulations under uniformisation can be carried out efficiently and whether there are any benefits of employing uniformisation methods.

To simulate a uniformised system one will introduce computational overhead by virtue of virtual reactions firing, and \textit{a priori} it might therefore be expected that uniformisation-based methods are outperformed by more standard approaches like Gillespie's DM. However, the performance benefits of approaches building on uniformisation have been observed in the literature, albeit mainly in the context of time-inhomogeneous reaction networks \cite{Voliotis2016}. We therefore, affirmatively, answer the question whether in the case of time-homogeneous systems, which is still the predominant modelling choice of practitioners, it is possible to find a stochastic simulation approach based on uniformisation which is as efficient as the aforementioned golden standard, Gillespie's DM. By addressing the issues of efficiency and adaptiveness of the uniformised systems we arrived at \Cref{algo:adaptive-IUDM} which, complexity-wise, is on par with Gillespie's DM. We reiterate that many improvements that have been made to the basic implementation of Gillespie's DM are equally applicable to the uniformised approaches discussed in this work, because the approach to choose the reaction channel that is to fire next is the same as for Gillespie's DM. The work in this paper therefore opens up the exploration of new approaches built on the uniformised systems by removing a previously perceived complexity problem. 

Extensions of the standard Gillespie's DM for time-homogeneous models to include delays have been proposed previously \cite{Barrio2006,Cai2007,Thanh2014,Thanh2017}. It is also possible to apply uniformisation to systems including delay reactions. In order to simulate such uniformised systems a first approach would be to explicitly generate the reaction times of the reactions as done for the time-inhomogeneous case, see \Cref{ap:time-inhomogeneous}. Thereafter we can proceed as in the DSSA algorithm in \cite{Barrio2006}. We note that this approach also explicitly generates the reaction times of the virtual reactions and therefore has an extra overhead cost relative to the original system. As an additional extension and application of uniformisation we note that one can apply tractable likelihood-based Bayesian inference techniques by using the uniformised system, e.g.\ \cite{Rao2013}. We leave exploration of complexity, delays, inference and uniformisation for future work. 

As mentioned in \Cref{sec:stratification}, one can use stratification to improve estimation of summary statistics for a wide variety of functionals $f$ of the sample path solutions. In the examples of \Cref{sec:stratification} we focused on the mean species numbers at a fixed time $T$, but reiterate that the stratification method is not restricted to such use cases only. Following the arguments from \Cref{sec:stratification} it is clear that for a general class of summary statistics we can expect a performance at least as good as implementations with Gillespie's DM. At the same time, however, we note that when applying stratification to estimate the one-dimensional marginal distributions of the species in the MAPK-cascade, \Cref{sec:mapk}, a significantly smaller variance reduction, as defined in terms of the mean integrated squared error, will be observed than when looking at the mean number of molecules as the summary statistic of interest. Another issue appears when one is interested in using a summary statistic that inherently depends on lower order moments of the random variable $\vect{Y}$, such as the population variance. To use (unbiased) stratified sampling for such summary statistics is a non-trivial task and a variance reduction is no longer guaranteed, see for example \cite{Wakimoto1970}. This raises the question as to what kind of problems, both in terms of the chemical reaction network and in terms of the employed summary statistic, result in a substantial variance reduction when combining uniformisation and stratification. We aim to investigate this in further work.

Finally, in this work we were primarily concerned with summary statistics that depend on the value of the paths at some time $T$, a commonly encountered choice of summary statistic. However, one might also be interested in path-dependent summary statistics, such as the transient evolution or mean value of a species over a series of time points $T_1,\dots,T_n$. A straightforward approach to generate such summary statistics with the standard SSAs would be to run the paths over $[0,T_1]$ and record the state, then over $[T_1,T_2]$ and record the state, etc. Such an approach applies equally well to the (improved) UDM, where we can either generate the total number of reactions $M$ on each interval individually or use the fact that given that $M$ reactions occur in $[0,T]$ the number of reactions in the intervals defined by $T_1,\dots,T_n$ follow a multinomial distribution. The efficient combination of transient analysis and uniformisation based methods is a subject of future work.

\subsection*{Acknowledgements}The authors would like to thank the anonymous reviewers, whose detailed comments improved the clarity of this paper. Casper H.\ L.\ Beentjes acknowledges the Clarendon Fund and New College, Oxford, for funding. Ruth E.\ Baker is a Royal Society Wolfson Research Merit Award holder and a Leverhulme Research Fellow, and also acknowledges the BBSRC for funding via grant no.\ BB/R000816/1. 
\newpage
\appendix

\section{Time-inhomogeneous models}\label{ap:time-inhomogeneous}
The results presented in this paper are valid for time-homogeneous Markov chains, i.e.\ models where the propensity functions do not have an inherent time-dependency and only depend on the state of the system $\vect{X}$. This assumption is generally sufficient to model intrinsic noise in a system, but whenever extrinsic noise effects are to be taken into account one has to relax this assumption. Extrinsic noise can be used to model the influence of external environments on the model behaviour and has been shown to influence the model dynamics \cite{Swain2002,Hilfinger2011,Bowsher2012,Voliotis2016,}.

\subsection{Stochastic model and simulation}
The most common modelling change to account for extrinsic noise is to assume that the reaction rate constants are allowed to be time varying functions, i.e.\ $c_k(t)$. This dependency can be prescribed as a simple function, e.g.\ $c_k(t) = 10(1+\sin(t))$, or be more elaborate, e.g.\ rates of the form $c_k(t) = 2\exp(Z(t))$ with $Z(t)$ some other stochastic process, which can, for example, be described by a stochastic differential equation (SDE). The result is a time-inhomogeneous Markov chain model and currently the best-known SSA for such systems is the Extrande method \cite{Voliotis2016} which uses the uniformisation method to avoid having to (numerically) integrate the propensity functions, $c_k(t)$. Methods built on the (numerical) integration of the propensity functions such as the modified next-reaction method \cite{Anderson2007} can also be constructed, but these methods do not compare favourably with methods based on uniformisation in general \cite{Voliotis2016}. The difference between the Extrande method and the UDM described in \Cref{sec:uniformisation} is that the former method, like the DM, generates all the reaction times in a serial fashion. In contrast to the case of time-homogeneous models, where it is not necessary to generate the reaction times explicitly for the UDM, in the case of a time-inhomogeneous model one has to know the reaction times to account for the explicit time dependence of the reaction propensities. We therefore have to adapt the UDM (\Cref{algo:GillespieUniform}) to account for this. As mentioned in \Cref{sec:uniformisation} we can sample the reaction times, $t_i$, conditional on the knowledge that the number of reactions firing in $[0,T)$ is $M$. We note that the reaction times $t_1,\dots,t_M$ are then distributed as the order statistics of $M$ uniform random variables in $[0,T)$. To generate these one can simply generate $M$ uniform random variables in $[0,T)$ and sort them in ascending order. Alternative methods to generate sorted uniform random numbers exist with better complexity properties, e.g.\ \cite[Chapter 3]{Devroye1986}. This extra step can be done prior to the reaction dynamics loop as depicted in \Cref{algo:GillespieUniformInhomogeneous}.

\begin{algorithm}[!ht]
\caption{Uniformised direct method for time-inhomogeneous models}
\label{algo:GillespieUniformInhomogeneous}
\begin{algorithmic}[1]
\Require{Initial data $\vect{X}_0$}
\Require{Stoichiometric matrix $\vect{\zeta}$}
\Require{Propensity functions $a_k(\vect{x},t)$}
\Require{Uniformisation rate $\bar{a}$}
\Require{Final time $T$}

\State{Generate $M\sim\mathcal{P}(\bar{a}T)$} \Comment{Total number of reactions that fire in $[0,T)$.}
\State{Generate $t_1,\dots,t_M\sim\mathcal{U}\left(0,T\right)$ s.t.\ $t_1<t_2<\dots<t_M$} \Comment{Reaction times.}
\For{$m=1,\dots,M$}
\State{Generate $u_1 \sim \mathcal{U}(0,1)$}
\Let{$\vect{A}_k$}{$a_k(\vect{X},t_m)$} \Comment{Calculate real reaction propensities.}
\Let{$\vect{A}_{K+1}$}{$\bar{a}-\sum_{k=1}^K \vect{A}_k$} \Comment{Calculate virtual reaction propensity.}
\State{Find $p$ such that $\sum_{k=1}^{p-1}\vect{A}_k<\bar{a} u_1\leq \sum_{k=1}^{p}\vect{A}_k$} \Comment{Choose next reaction to fire.}
\If {$p\in\{1,\dots,K\}$} \Comment{Only need to fire real reactions.}
\Let{$\vect{X}$}{$\vect{X}+\vect{\zeta}_p$} \Comment{Update state vector.}
\EndIf
\EndFor
\end{algorithmic}
\end{algorithm}

We note that if $a_0>\bar{a}$ is observed in the course of a sample path the procedure to adapt the uniformisation rate is the same as described in \Cref{sec:adaptive-uniformisation} apart from the fact that the time $T^*$ does not have to be sampled in this case, we can simply set $T^*=t_{M^*}$ with $M^*$ the number of reactions that have fired up until the point at which $a_0>\bar{a}$.

The extra computational gain that was achieved by firing virtual reactions consecutively does not apply in this framework. For time-homogeneous models we used the fact that the distribution of the number of consecutive virtual reactions firing was geometric, but this is no longer is valid for time-inhomogeneous models. Due to the explicit dependence on time of the propensity functions we get an expression for the distribution of the number of consecutive virtual reactions of the form

\begin{multline}
\mathbb{P}\left(r \text{ consecutive virtual reactions before next real reaction fires} \right) =\\ \frac{a_0(t_{r+1})}{\bar{a}}\prod_{i=1}^{r} \left(1-\frac{a_0(t_i)}{\bar{a}}\right).
\end{multline}

This is no longer a single parameter distribution and, although it is analytically tractable, sampling from it will generally involve evaluating the total propensity $a_0$ at the time points of the virtual reactions. Therefore such a strategy might potentially negate any speed-ups from generating consecutive virtual reactions in a single step.

As a result we do not want to have to choose a uniformisation rate $\bar{a}$ that is much larger than $a_0$ over $[0,T)$ as every virtual reaction will have to be simulated individually for the time-inhomogeneous case in \Cref{algo:GillespieUniformInhomogeneous}. If the total propensity $a_0$ varies strongly over the course of $[0,T)$ it might be beneficial to regularly adapt the rate $\bar{a}$, which in the extreme case of adapting the rate after every reaction yields the Extrande method.

\subsection{Stratification}
The theory for stratification still holds in the case of time-in\-ho\-mo\-ge\-neous models. If one can therefore feasibly generate sample paths using \Cref{algo:GillespieUniformInhomogeneous} with a fixed uniformisation rate, $\bar{a}$, it would be wise to apply stratification with respect to the number of reactions firing, $M$, to achieve a variance reduction at no extra cost. If, however, the propensity $a_0$ changes noticeably over $[0,T)$ one might want to use an adaptive uniformisation rate, $\bar{a}$, or the Extrande method, in which case the stratification method is no longer valid, because the reaction times in the interval $[0,T)$ are no longer drawn from a single parameter joint distribution that is the same for each sample path.

\section{Unbiased adapting of the uniformisation rate}\label{ap:adapting}
Here we provide an alternative view on the sampling of the time $T^*$ at which one adapts the uniformisation rate, as discussed in \Cref{sec:adaptive-uniformisation}. At the same time we show that the conditional sampling of $T^*$ does in fact yield unbiased samples from the correct distribution.

\subsection{Derivation of the conditional distribution of $T^*$}
For completeness we first show that the $M^*$-th order statistic of $M$ uniform random variables follows a Beta distribution. We start with the assumption that we know that $M$ reactions fire in the time interval $[0,T)$ as sampled from the Poisson distribution with parameter $\bar{a}T$. Their reaction times are then uniformly distributed on $[0,T)$. To find the distribution of the time $T^*$ of the $M^*$-th reaction we note that in order to have $T^*\in [t,t+\Delta t)$ we need exactly $M^*-1$ reactions in $[0,t)$, one reaction in $[t,t+\Delta t)$ and $(M-M^*)$ reactions in $[t+\Delta t,T)$. This yields the following expression
\begin{equation}
\mathbb{P}\left(T^*\in [t,t+\Delta t)\right)=\frac{M!}{(M-M^*)!(M^*-1)!}\left(\frac{t}{T}\right)^{M^*-1}\frac{\Delta t}{T}\left(1-\frac{t+\Delta t}{T} \right)^{M-M^*},
\end{equation}
where the pre-factor stems from the indistinguishable nature of the reactions in $[0,t)$ and $[t+\Delta t,T)$. By letting $\Delta t\to 0$ we get the probability density function $f(t)$ for $T^*$
\begin{equation}
f(t) = \frac{\Gamma(M-1)}{\Gamma(M-M^*-1)\Gamma(M^*)}\left(\frac{t}{T}\right)^{M^*-1}\frac{1}{T}\left(1-\frac{t}{T} \right)^{M-M^*}.
\end{equation}
We now let $u=t/T$, i.e.\ we consider the distribution of $T^*/T$, and recognize the pre-factor as $1/B(M-M^*+1,M^*)$, where $B$ is the beta-function. This leads to the following probability density function $g(u)$ for $T^*/T$
\begin{equation}
g(u)=\frac{1}{B(M-M^*+1,M^*)}\left(u\right)^{M^*-1}\left(1-u \right)^{M-M^*},
\end{equation}
which is readily seen to be equivalent to the Beta distribution on $[0,1)$ with parameters $M^*$ and $(M-M^*)+1$.

\subsection{Unconditional distribution of $T^*$}
In our approach to adapt the uniformisation rate we sample a time $T^*$ conditional on the fact that $M$ reactions within the interval $[0,T)$ were sampled at the uniformisation rate $\bar{a}$. For the unconditional distribution of $T^*$, the time of the $M^*$-th reaction and the point at which we adapt the uniformisation rate, as in \Cref{algo:adaptive-IUDM}, we start with the following observation for $t\in[0,T)$
\begin{align*}
\mathbb{P}\left(T^*\leq t)\right)&=\sum_{M=0}^{\infty} \mathbb{P}\left(T^*\leq t|M\text{ reactions in }[0,T)\right)\mathbb{P}\left(M\text{ reactions in }[0,T)\right)\\
&=\sum_{M=M^*}^{\infty} \mathbb{P}\left(\frac{T^*}{T}\leq \frac{t}{T} \middle| M\text{ reactions in }[0,T)\right)\mathbb{P}\left(M\text{ reactions in }[0,T)\right),
\end{align*}
where we now recognize the first term in the sum to be described by a Beta distribution (with parameters $M^*$ and $(M-M^*)+1$) and the second term by a Poisson distribution (with parameter $\bar{a}T$). Substitution of the relevant expressions for these distributions then yields
\begin{align*}
\mathbb{P}\left(T^*\leq t)\right)&=\sum_{M=M^*}^{\infty} \left(\frac{1}{B(M^*,(M-M^*)+1)}\int_{0}^{t/T}s^{M^*-1}(1-s)^{M-M^*}\id s\right)\frac{\left(\bar{a}T\right)^M}{M!}e^{-\bar{a}T}\\
&=\frac{1}{\Gamma(M^*)} \int_{0}^{t/T} s^{M^*-1} \left(\bar{a}T\right)^{M^*}e^{-\bar{a}Ts} \left(\sum_{M=M^*}^{\infty} \frac{\left(\bar{a}T(1-s)\right)^{M-M^*} }{(M-M^*)!}e^{-\bar{a}T(1-s)} \right)\id s\\
&=\frac{1}{\Gamma(M^*)} \int_{0}^{\bar{a}t} u^{M^*-1} e^{-u} \id u,
\end{align*}
which is the distribution function for the Gamma$(M^*,\bar{a})$ distribution. This is exactly the distribution of $M^*$ i.i.d.\ exponential random variables with parameter $\bar{a}$. As a result $T^*$ is distributed as the $M^*$-th reaction time in a system with constant propensity $\bar{a}$. This proves that the conditional sampling of $T^*$ as described in \Cref{sec:adaptive-uniformisation} yields in fact samples from the correct distribution and therefore makes the construction of sample paths with an adaptive uniformisation rate unbiased.

\setlength\bibitemsep{4pt}
\printbibliography
\end{document}